\PassOptionsToPackage{table}{xcolor}

\documentclass[sigconf]{acmart}

\AtBeginDocument{%
	\providecommand\BibTeX{{%
			\normalfont B\kern-0.5em{\scshape i\kern-0.25em b}\kern-0.8em\TeX}}}

\usepackage[T1]{fontenc}            
\usepackage[utf8]{inputenc}         
\usepackage{microtype}              
\usepackage[scaled=.8]{beramono}    
\usepackage[abbreviations]{foreign} 
\usepackage{balance}                
\usepackage{array}
\usepackage{booktabs}
\usepackage{graphicx}
\usepackage{multirow}
\usepackage{pifont}
\usepackage{tcolorbox}
\usepackage{xcolor}
\usepackage{colortbl}
\usepackage{subcaption}
\usepackage{xfp}
\usepackage[all]{nowidow}
\usepackage{cleveref} 

\hypersetup{colorlinks=true,allcolors=blue!80}

\newcommand{\what}[0]{\emph{What}\xspace}
\newcommand{\why}[0]{\emph{Why}\xspace}
\newcommand{\fix}[0]{\emph{Fix}\xspace}
\newcommand{\txt}[0]{\emph{Text}\xspace}
\newcommand{\source}[0]{\emph{Code}\xspace}
\newcommand{\link}[0]{\emph{Hyperlink}\xspace}
\newcommand{\links}[0]{\emph{Hyperlinks}\xspace}

\newtcolorbox{finding}[1][]{%
  colback=blue!5,colframe=blue!20
  #1}

\newcommand{\tableCell}[1]{%
    \cellcolor[RGB]{\inteval{255-#1}, \inteval{255-#1}, 255}%
    \ifnum#1=0%
        \quad%
    \else%
        #1%
    \fi%
}

\copyrightyear{2024}
\acmYear{2024}
\setcopyright{none}\acmConference[ICPC '24]{32nd IEEE/ACM International Conference on Program Comprehension}{April 15--16, 2024}{Lisbon, Portugal}
\acmBooktitle{32nd IEEE/ACM International Conference on Program Comprehension (ICPC '24), April 15--16, 2024, Lisbon, Portugal}
\acmDOI{10.1145/3643916.3644404}
\acmISBN{979-8-4007-0586-1/24/04}

\begin{document}

\title{What the Fix? A Study of ASATs Rule Documentation}

\author{Corentin Latappy}
\email{corentin.latappy@labri.fr}
\affiliation{%
	\institution{Univ. Bordeaux, CNRS, Bordeaux INP, LaBRI, UMR 5800}
	\institution{Promyze}
	\country{France}
}

\author{Thomas Degueule}
\email{thomas.degueule@labri.fr}
\affiliation{%
	\institution{Univ. Bordeaux, CNRS, Bordeaux INP, LaBRI, UMR 5800}
	\country{France}
}

\author{Jean-Rémy Falleri}
\email{falleri@labri.fr}
\affiliation{%
	\institution{Univ. Bordeaux, CNRS, Bordeaux INP, LaBRI, UMR 5800}
	\institution{Institut Universitaire de France}
	\country{France}
}

\author{Romain Robbes}
\email{romain.robbes@labri.fr}
\affiliation{%
	\institution{Univ. Bordeaux, CNRS, Bordeaux INP, LaBRI, UMR 5800}
	\country{France}
}

\author{Xavier Blanc}
\email{xavier.blanc@labri.fr}
\affiliation{%
	\institution{Univ. Bordeaux, CNRS, Bordeaux INP, LaBRI, UMR 5800}
	\country{France}
}

\author{Cédric Teyton}
\email{cedric.teyton@promyze.com}
\affiliation{%
	\institution{Promyze}
	\country{France}
}

\begin{abstract}
Automatic Static Analysis Tools (ASATs) are widely used by software developers to diffuse and enforce coding practices. Yet, we know little about the documentation of ASATs, despite it being critical to learn about the coding practices in the first place. We shed light on this through several contributions. First, we analyze the documentation of more than 100 rules of 16 ASATs for multiple programming languages, and distill a taxonomy of the purposes of the documentation---\what triggers a rule; \why it is important; and how to \fix an issue---and its types of contents. Then, we conduct a survey to assess the effectiveness of the documentation in terms of its goals and types of content. We highlight opportunities for improvement in ASAT documentation. In particular, we find that the \why purpose is missing in half of the rules we survey; moreover, when the \why is present, it is more likely to have quality issues than the \what and the \fix.

\end{abstract}

\ccsdesc[500]{Software and its engineering~Software libraries and repositories}
\ccsdesc[500]{Software and its engineering~Software maintenance tools}
\ccsdesc[300]{Software and its engineering~Software design engineering}

\keywords{software quality, automatic static analysis tools, linters, documentation}


\maketitle

\section{Introduction}

Automatic Static Analysis Tools (ASATs, sometimes called linters)~\cite{novak_taxonomy_2010} are very popular quality insurance tools used in a quarter~\cite{tomasdottir_adoption_2020} to half of software projects~\cite{beller_analyzing_2016}. According to the \url{https://analysis-tools.dev} website, there are more than 600 ASATs.
The core principle behind these tools is to scan a code base looking for evidence of potential issues with the source code, such as not following best practices, using error-prone constructs, and detecting security or performance problems~\cite{novak_taxonomy_2010,mclean_comparing_2012,ashfaq_comparative_2019,habchi_adopting_2018,tahaei_security_2021}.
When such an issue is detected, a warning is issued to the developers of the project so that they can inspect the incriminated code and fix the issue if necessary.
Issues are generally defined by \emph{rules}, a popular example being the \texttt{eqeqeq} rule in ESLint JavaScript's ASAT that advises developers to use \texttt{===} instead of \texttt{==} to avoid type coercion errors.

Even though these tools are prone to issue false-positive warnings~\cite{kang_detecting_2022,kim_filtering_2010,jung_taming_2005} and their effect on software quality is still a matter of debate~\cite{tomassi_bugs_2018,tomassi_real-world_2021, kavaler_tool_2019}, one positive aspect raised by the use of ASATs is their ability to improve the developer's knowledge and skills and to support the onboarding of newcomers~\cite{tomasdottir_why_2017, do_why_2022}.
In this article, we focus on the following scenario of ASATs: a developer encounters a warning for the first time and wants to learn more about it.
Usually, warnings raised by ASATs within the IDE or within the terminal as a result of the use of its command line interface only contain limited information~\cite{smith_why_2020, tahaei_security_2021}.
To further explain a given rule, ASATs generally offer more extensive documentation via formatted documents available online.

We argue that this documentation is essential for improving developers' knowledge and skills, as well as promoting compliance with the rules.
Even though warning notifications have already been studied~\cite{smith_why_2020, tahaei_security_2021}, empirical evidence shows that their quality is not ideal~\cite{tahaei_security_2021,johnson_cross-tool_2016,nachtigall_large-scale_2022}.
Furthermore, to the best of our knowledge, there is no comprehensive study on the \textit{reference documentation} of ASATs rules, which developers look for after the initial warning.
We believe that the reference documentation of ASATs rules is not a classical piece of software documentation, such as an API documentation~\cite{watson_api_2013}.
Indeed, ASATs rule documentation is challenging as it needs to explain issues regardless of their particular context in such a manner that developers are able later to understand them in their own context.
Another difficult point is that ASATs rules are sometimes subjective, for instance those related to the use of goto statements~\cite{nagappan_empirical_2015} or the classical dilemma of choosing between snake and camel case~\cite{binkley_camelcase_2009}. 
Thus, little is known about how to best document ASAT rules. This lack of knowledge is damaging when ASATs such as Semgrep allow developers around the world to effortlessly create and share hundreds of rules.

In this article, we explore the current state of the ASATs' rules documentation and contrast it with developers' expectations, in order to extract best practices with the goal to increase the quality of this documentation. We start with an empirical study of how real-world ASAT rules are documented. \Cref{sec:analysis} details our study of more than 100 rules across 16 ASATs, covering 7 programming languages (plus two polyglot ASATs). Through an iterative analysis of these rules, we derive a nomenclature covering 15 documentation attributes in three themes. In \Cref{sec:taxo}, we distill our nomenclature in a simple taxonomy of documentation purposes (\what triggers the rule, \why it is important, and how to \fix the issue) and content types (\txt, \source, and \links). Only half of the rules we analyze have a \why purpose.

Then, we use our taxonomy to contrast the documentation of 12 real-world rules with developer expectations via a survey. \Cref{sec:survey} presents the survey in which 85 respondents evaluated the rules 298 times. Among other findings, developers highlight quality issues with the documentation of the \why purpose, and emphasize the pedagogical aspects and the need for conciseness in ASAT documentation. To close the paper, we discuss the limitations of this study (\Cref{sec:threats}), other studies of ASATs and documentation (\Cref{sec:rw}), before concluding (\Cref{sec:conclusion}).
A replication package including raw data, the nomenclature and taxonomy, and survey results is available online\footnote{\url{https://icpc2024-asats.github.io}} and on Zenodo~\cite{latappy_replication_2024}.

\section{A Nomenclature for Rule Documentation}
\label{sec:analysis}


The first step of our study is to create a nomenclature based on the documentation of the rules provided by ASATs.
A nomenclature is a classification tool that provides a structured and systematic way of naming and referring to a wide range of objects. This nomenclature will help us define the types of information we find in a rule's documentation and group them by purpose. To build it, we use a three-step process described next: (1) we select a total of 16 diverse ASATs that provide documentation for their rules; (2) we iteratively build a corpus of 119 rules, for which we code the documentation concepts we encounter to support their comparison; and (3) we compare and analyze the rule documentations to provide the expected nomenclature, which covers 15 different concepts.

\subsection{Selecting ASATs}

A GitHub project listing ASATs\footnote{\url{https://github.com/analysis-tools-dev/static-analysis}, 12K stars} references over 600 different tools; choosing a reasonable subset is both necessary and challenging.
A first hard requirement is that we only include ASATs focusing on code rule enforcement, leaving out for instance code beautifiers.
A second hard requirement is that we only include ASATs that provide a website presenting the rules documentation.
Finally, we select ASATs emphasizing both \emph{diversity} and \emph{popularity}.
Since we want our nomenclature to be used beyond our study, regardless of ASAT, we include a diverse set of ASATs targetting different programming languages.
Finally, we want to bias our corpus toward popular ASATs as we postulate that popular ASATs have more odds of including well-thought-out and comprehensive documentation.

We define two criteria to judge diversity and one criterion to judge popularity.
The two diversity criteria we propose are: that the ASATs covers most of the popular programming languages\footnote{\url{https://madnight.github.io/githut/\#/pull_requests/2021/4}}, and to have at least two ASATs per programming language covered; we also include ASAT that supports multiple languages to account for this category.
The popularity criterion we propose is that ASATs must have a GitHub repository with at least 1000 stars.

Starting with the ASAT list, we select the 16 ASATs in \cref{tab:lang_and_asat} using our criteria and the domain expertise of our industrial partner.
The only exception to the popularity criterion is Gendarme, which we pick to increase diversity: it is the official ASAT for the alternative Mono C\# implementation.
We are aware that this selection is somewhat arbitrary and that another selection could have been made with the same criteria; we discuss this issue in~\cref{sec:threats}.

\begin{table}
    \small
    \centering
    \caption{Languages and ASAT selected}
    \begin{tabular}{@{}rl|rl@{}}
        \toprule
            Languages                & ASATs                                                                             & Languages               & ASATs                                                             \\
        \midrule
            \multirow{2}{*}{C / C++} & \href{https://oclint.org}{OCLint}                                                 & \multirow{2}{*}{PHP}    & \href{https://github.com/FriendsOfPHP/PHP-CS-Fixer}{PHP CS Fixer} \\
                                     & \href{https://cppcheck.sourceforge.io}{Cppcheck}                                  &                         & \href{https://psalm.dev}{Psalm}                                 \\
        \addlinespace
            \multirow{2}{*}{C\#}     & \href{https://www.mono-project.com/docs/tools+libraries/tools/gendarme}{Gendarme} & \multirow{2}{*}{Python} & \href{https://pylint.pycqa.org/en/latest/intro.html}{Pylint}      \\
                                     & \href{https://github.com/dotnet/roslynator}{Roslynator}                                         &                         & \href{https://www.flake8rules.com}{Flake8}                        \\
        \addlinespace
            \multirow{2}{*}{Java}    & \href{https://checkstyle.org}{Checkstyle}                                         & \multirow{2}{*}{Ruby}   & \href{https://docs.rubocop.org/rubocop}{RuboCop}                  \\
                                     & \href{http://findbugs.sourceforge.net/index.html}{SpotBugs}                       &                         & \href{https://brakemanscanner.org}{Brakeman}                      \\
        \addlinespace
            \multirow{2}{*}{JS / TS} & \href{https://eslint.org}{ESLint}                                                 & \multirow{2}{*}{Multi}  & \href{https://semgrep.dev}{Semgrep}                               \\
                                     & \href{https://rslint.org}{RSLint}                                                 &                         & \href{https://www.sonarlint.org}{SonarLint}                       \\
        \bottomrule
    \end{tabular}
    \label{tab:lang_and_asat}
\end{table}


\subsection{Coding Documentation Concepts}

Our selection of ASATs offer rule documentation that differs in structure and in content, and this can even happen between two rules from the same ASAT. To compare ASAT documentations, we identify and align all the concepts used in the documentation. 

\begin{figure}
\includegraphics[width=\linewidth]{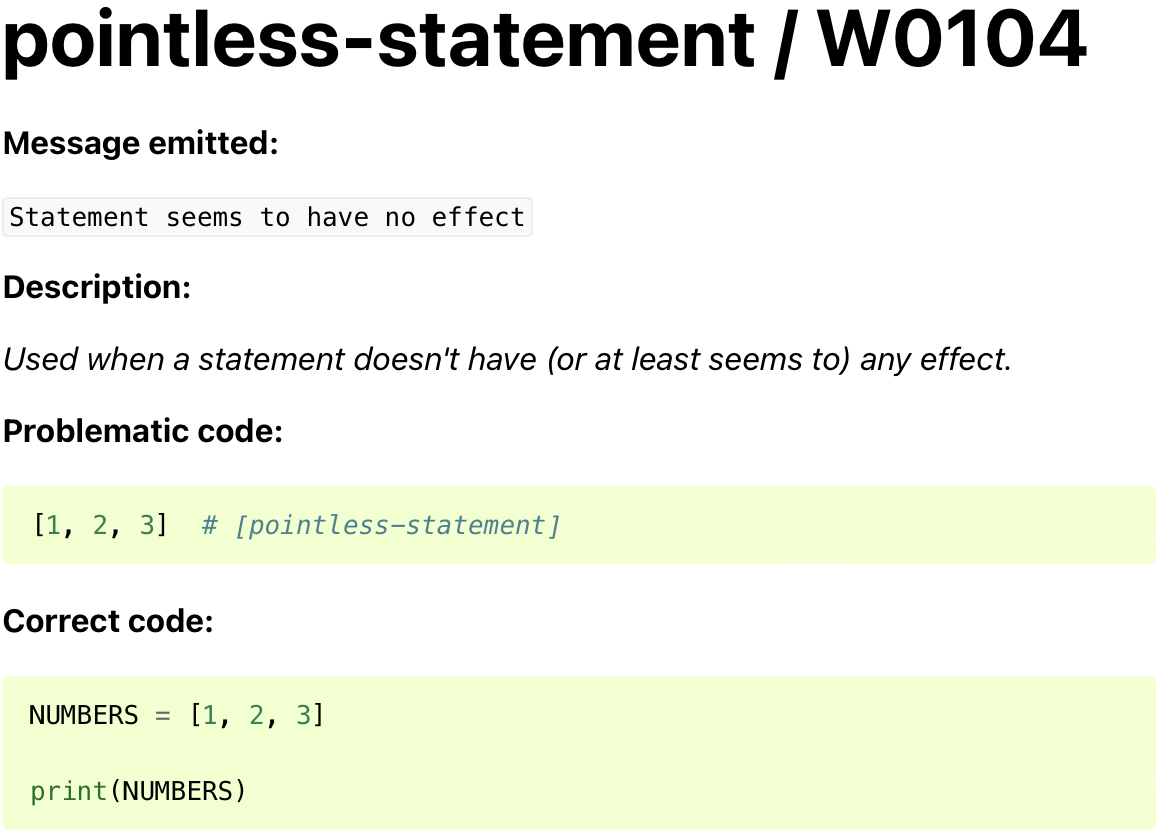}
  \caption{Documentation of the rule \textit{pointless-statement} from \texttt{Pylint}}
\label{fig:pylint_rule_documentation}
\end{figure}

For example,~\cref{fig:pylint_rule_documentation} displays the documentation for the rule \textit{pointless-statement} provided by \texttt{Pylint} . 
Using visual and graphical cues on this documentation (headings, line breaks, boxes, images, etc.), we extract and code four concepts: an \emph{emitted message} (error message when the rule is violated), a \emph{description}, a \emph{problematic code}, and a \emph{correct code}.
If we now want to compare this documentation with another rule, we need to align their documentation concepts.
This alignment will, e.g., reveal which concepts are present in both documentations, or which ones are present in only one rule.

Each ASATs provides many rules. We use a two-step process to first code and then reconcile the documentation concepts. More precisely, we build an independent coding for each ASAT iteratively and then reconcile them into a single and global coding for rules documentation.
For each ASAT, we pick up rules at random, look at the documentation for visual and graphical cues, and add documentation concepts if they were not identified before. We repeat this until saturation---analyzing 5 successive rules without discovering a new concept. Using this process, we analyze at least 6 rules per ASAT;~\cref{tab:rules_extracted}, column 3, counts the rules we inspect per ASAT. 

The reconciliation of the documentation concepts identified in each ASAT was done manually by the authors during a harmonization session.
For instance, we consider that the \textbf{Message~emitted} concept in \texttt{Pylint} rule of~\cref{fig:pylint_rule_documentation} is highly similar to the \textbf{Message~output} concept identified in \texttt{Cppcheck} rules. We therefore reconcile these two concepts and code them with \textbf{Error~Output}.

As a result, we build a corpus of 119 rules, with which we identify 15 documentation concepts across ASATs (see~\cref{tab:rules_extracted}). Documentation consistency vary: depending on the tool, we need between 6 and 10 rules to reach saturation.

\begin{table*}
    \small
    \centering
    \caption{Percentage of presence of documentation concepts for each ASAT}
    \begin{tabular}{rlcccccccccccccccc}
        \toprule
                                        &              &                          & \multicolumn{4}{c}{\textbf{Comprehension}}                                                                                                     & \multicolumn{7}{c}{\textbf{Usage}}                                                                                                                                                                                  & \multicolumn{4}{c}{\textbf{Metadata}} \\
            \cmidrule(r){4-7} \cmidrule(lr){8-14} \cmidrule(l){15-18}
            \textbf{Language}                    & \textbf{ASAT}         & \textbf{\# Rules} & \rotatebox{80}{Code Example} & \rotatebox{80}{Description} & \rotatebox{80}{Further Information} & \rotatebox{80}{When Not To Use It} & \rotatebox{80}{Auto Fix} & \rotatebox{80}{Compatibility} & \rotatebox{80}{Configurations} & \rotatebox{80}{Error Output} & \rotatebox{80}{IDE Fix} & \rotatebox{80}{Since} & \rotatebox{80}{Usage Example} & \rotatebox{80}{Related Rules} & \rotatebox{80}{Rule Definition} & \rotatebox{80}{Rule Set} & \rotatebox{80}{Severity} \\
        \midrule
            \multirow{2}{*}{C++}        & OCLint       & 6                        & \tableCell{100}              & \tableCell{100}             & \tableCell{0}                       & \tableCell{0}                      & \tableCell{0}            & \tableCell{0}                 & \tableCell{0}                  & \tableCell{0}                & \tableCell{0}           & \tableCell{100}       & \tableCell{0}                 & \tableCell{0}                 & \tableCell{100}                 & \tableCell{0}            & \tableCell{0}            \\
                                        & Cppcheck     & 6                        & \tableCell{100}              & \tableCell{100}             & \tableCell{83}                      & \tableCell{0}                      & \tableCell{0}            & \tableCell{0}                 & \tableCell{0}                  & \tableCell{67}               & \tableCell{0}           & \tableCell{0}         & \tableCell{0}                 & \tableCell{100}               & \tableCell{0}                   & \tableCell{0}            & \tableCell{100}          \\
            \multirow{2}{*}{C\#}         & Gendarme     & 9                        & \tableCell{89}               & \tableCell{100}             & \tableCell{0}                       & \tableCell{0}                      & \tableCell{0}            & \tableCell{22}                & \tableCell{11}                 & \tableCell{0}                & \tableCell{0}           & \tableCell{33}        & \tableCell{0}                 & \tableCell{0}                 & \tableCell{0}                   & \tableCell{0}            & \tableCell{0}            \\
                                        & Roslynator   & 10                       & \tableCell{100}              & \tableCell{0}               & \tableCell{10}                      & \tableCell{0}                      & \tableCell{0}            & \tableCell{0}                 & \tableCell{10}                 & \tableCell{0}                & \tableCell{0}           & \tableCell{0}         & \tableCell{0}                 & \tableCell{0}                 & \tableCell{0}                   & \tableCell{0}            & \tableCell{100}          \\
            \multirow{2}{*}{Java}       & Checkstyle   & 6                        & \tableCell{100}              & \tableCell{100}             & \tableCell{67}                      & \tableCell{0}                      & \tableCell{0}            & \tableCell{0}                 & \tableCell{83}                 & \tableCell{100}              & \tableCell{0}           & \tableCell{100}       & \tableCell{100}               & \tableCell{0}                 & \tableCell{0}                   & \tableCell{100}          & \tableCell{0}            \\
                                        & SpotBugs     & 6                        & \tableCell{0}                & \tableCell{100}             & \tableCell{33}                      & \tableCell{0}                      & \tableCell{0}            & \tableCell{0}                 & \tableCell{0}                  & \tableCell{0}                & \tableCell{0}           & \tableCell{0}         & \tableCell{0}                 & \tableCell{0}                 & \tableCell{0}                   & \tableCell{0}            & \tableCell{0}            \\
            \multirow{2}{*}{JS} & ESLint       & 10                       & \tableCell{70}               & \tableCell{100}             & \tableCell{40}                      & \tableCell{50}                     & \tableCell{40}           & \tableCell{20}                & \tableCell{80}                 & \tableCell{0}                & \tableCell{20}          & \tableCell{100}       & \tableCell{0}                 & \tableCell{30}                & \tableCell{100}                 & \tableCell{0}            & \tableCell{0}            \\
                                        & RSLint       & 8                        & \tableCell{100}              & \tableCell{100}             & \tableCell{0}                       & \tableCell{0}                      & \tableCell{0}            & \tableCell{0}                 & \tableCell{38}                 & \tableCell{0}                & \tableCell{0}           & \tableCell{0}         & \tableCell{0}                 & \tableCell{0}                 & \tableCell{100}                 & \tableCell{0}            & \tableCell{0}            \\
            \multirow{2}{*}{PHP}        & PHP CS Fixer & 7                        & \tableCell{100}              & \tableCell{100}             & \tableCell{0}                       & \tableCell{29}                     & \tableCell{100}          & \tableCell{0}                 & \tableCell{29}                 & \tableCell{0}                & \tableCell{0}           & \tableCell{0}         & \tableCell{0}                 & \tableCell{0}                 & \tableCell{0}                   & \tableCell{100}          & \tableCell{29}           \\
                                        & Psalm        & 7                        & \tableCell{100}              & \tableCell{100}             & \tableCell{14}                      & \tableCell{0}                      & \tableCell{0}            & \tableCell{0}                 & \tableCell{14}                 & \tableCell{0}                & \tableCell{0}           & \tableCell{0}         & \tableCell{0}                 & \tableCell{0}                 & \tableCell{0}                   & \tableCell{0}            & \tableCell{0}            \\
            \multirow{2}{*}{Python}     & Pylint       & 10                       & \tableCell{100}              & \tableCell{100}             & \tableCell{10}                      & \tableCell{0}                      & \tableCell{0}            & \tableCell{0}                 & \tableCell{0}                  & \tableCell{100}              & \tableCell{0}           & \tableCell{0}         & \tableCell{0}                 & \tableCell{0}                 & \tableCell{0}                   & \tableCell{0}            & \tableCell{100}          \\
                                        & Flake8       & 6                        & \tableCell{100}              & \tableCell{100}             & \tableCell{83}                      & \tableCell{0}                      & \tableCell{0}            & \tableCell{0}                 & \tableCell{0}                  & \tableCell{0}                & \tableCell{0}           & \tableCell{0}         & \tableCell{0}                 & \tableCell{0}                 & \tableCell{0}                   & \tableCell{0}            & \tableCell{0}            \\
            \multirow{2}{*}{Ruby}       & RuboCop      & 9                        & \tableCell{89}               & \tableCell{100}             & \tableCell{56}                      & \tableCell{0}                      & \tableCell{78}           & \tableCell{0}                 & \tableCell{56}                 & \tableCell{0}                & \tableCell{0}           & \tableCell{100}       & \tableCell{0}                 & \tableCell{0}                 & \tableCell{0}                   & \tableCell{0}            & \tableCell{0}            \\
                                        & Brakeman     & 7                        & \tableCell{71}               & \tableCell{100}             & \tableCell{57}                      & \tableCell{0}                      & \tableCell{0}            & \tableCell{0}                 & \tableCell{0}                  & \tableCell{43}               & \tableCell{0}           & \tableCell{0}         & \tableCell{0}                 & \tableCell{0}                 & \tableCell{0}                   & \tableCell{0}            & \tableCell{0}            \\
            \multirow{2}{*}{Multi}      & Semgrep      & 6                        & \tableCell{100}              & \tableCell{100}             & \tableCell{50}                      & \tableCell{0}                      & \tableCell{0}            & \tableCell{0}                 & \tableCell{0}                  & \tableCell{0}                & \tableCell{0}           & \tableCell{0}         & \tableCell{0}                 & \tableCell{0}                 & \tableCell{100}                 & \tableCell{0}            & \tableCell{100}          \\
                                        & SonarLint    & 6                        & \tableCell{100}              & \tableCell{100}             & \tableCell{17}                      & \tableCell{0}                      & \tableCell{0}            & \tableCell{0}                 & \tableCell{0}                  & \tableCell{0}                & \tableCell{0}           & \tableCell{0}         & \tableCell{0}                 & \tableCell{0}                 & \tableCell{0}                   & \tableCell{0}            & \tableCell{100}          \\
        \midrule
            & Total & 119 & \tableCell{87} & \tableCell{92} & \tableCell{30} & \tableCell{6} & \tableCell{15} & \tableCell{3} & \tableCell{22} & \tableCell{19} & \tableCell{2} & \tableCell{29} & \tableCell{5} & \tableCell{8} & \tableCell{25} & \tableCell{11} & \tableCell{34} \\
        \bottomrule
    \end{tabular}
    \label{tab:rules_extracted}
\end{table*}

\subsection{The Nomenclature}

After identifying documentation elements in the previous phase, to finalize our nomenclature, we grouped the documentation concepts we encountered into different themes to clarify their role. This classification was made by the authors following an approach similar to open card sorting~\cite{zimmermann_card-sorting_2016}. We obtained the following three themes:
\begin{itemize}
    \item \textbf{Comprehension}, which contains all the concepts identifying parts of the documentation that help to understand the rule.
    \item \textbf{Usage}, which contains all concepts identifying parts explaining how to correctly configure the rule for a given project.
    \item \textbf{Metadata}, which contains all the concepts identifying ASAT-specific information (such as organizational scheme or rules source code).
\end{itemize}

Our nomenclature is presented in~\cref{tab:rules_extracted} with 15 documentation concepts grouped into three themes. For the sake of readability, we present the results by ASAT, even if there are differences between rules within the same ASAT.
For each ASAT and concept,~\cref{tab:rules_extracted} shows the percentage of the concept's presence across all analyzed rules (the definition of each concept and the full per-rule version are available in our replication kit\footnote{\url{https://icpc2024-asats.github.io?page=analysis&tab=nomenclature}}).

The first point of interest is that \emph{there is no single documentation concept that can be found in all ASATs}.
The two most common concepts are \emph{description} (100\% in all but one ASAT, \texttt{Roslynator}) and \emph{code example} (70--100\% in all but one ASAT, \texttt{SpotBugs}). This presence is mirrored in the rules: 109 out of 119 rules have descriptions, and 104 out of 119 have code examples. We were surprised by the lack of descriptions for \texttt{Roslynator}, especially since every other ASAT had a description: for this ASAT, the title of the rule acts as a description \footnote{e.g., \url{https://josefpihrt.github.io/docs/roslynator/analyzers/RCS0033}}.
Other concepts are sparser, with 40 rules out of 119 (6 ASATs) that include a \emph{severity}, and 36 out of 119 that include \emph{further information} (12 ASATs). Some concepts are very sparse, such as \emph{IDE Fix} (2 rules out of 119) and \emph{compatibility} (4 out of 119).

The second point of interest is that, apart from the description, the code example, and the further information concepts, \emph{few ASATs share the same concepts}. We note that the average number of concepts per ASAT is 5.1, with a median of 4.5.
\texttt{ESLint} is the exception, with 11 concepts used, albeit with only three concepts used consistently (\emph{description}, \emph{since}, and \emph{rule definition}). 

The final point of interest is that few ASATs are consistent in their use of the documentation elements. Only \texttt{OCLint} is completely consistent, while \texttt{ESLint} is the most inconsistent.

\begin{finding}
\textbf{Finding \#1:} Projecting our nomenclature onto the rules of the 16 ASATs we selected clearly reveals the differences in terms of rule documentation. This reinforces the need to define a more abstract taxonomy and to carry out a survey of developers to better understand their expectations and calls for action, which is the purpose of the following sections. 
\end{finding}
\section{Taxonomy of Content Purposes and Types}
\label{sec:taxo}

In this section, we dive deeper into the actual content of rules documentation to better understand their purpose.
We focus on the nomenclature's \textbf{Comprehension} theme as, we are interested in the developer's comprehension of ASAT rules and the best practices they describe.
In contrast, the nomenclature's \textbf{Usage} theme relates to the usage of the ASAT (.e.g, how to configure a rule), while \textbf{Metadata} mainly contains ASAT-specific information (such as categories of rules or source code).

The \textbf{Comprehension} theme consists of four terms:~\emph{Code Example}, \emph{Description}, \emph{Further Information}, and \emph{When Not To Use It} (\Cref{tab:rules_extracted}); we focus on this theme in the following.
When looking at the data, one quickly realizes that these umbrella terms conceal a rich underlying diversity of purposes and content types.
For instance, \emph{Description} content sometimes highlights the rationale for a particular rule; at other times, it describes how to identify code that breaches the rule. Similarly, \emph{Code Example} snippets may present compliant code, non-compliant code, or a mix thereof. The content also uses a combination of text, hyperlinks, and source code.

\Cref{sec:taxo-protocol} presents the methodology we follow to extract and consolidate this information; \Cref{sec:taxo-validation} describe how we validate the resulting taxonomy internally (the survey in \Cref{sec:survey} provides additional validation). Finally, \Cref{sec:taxo-results} presents the results of applying our taxonomy to the rules of \Cref{tab:rules_extracted}.

\subsection{Extraction}
\label{sec:taxo-protocol}

To better understand what is the purpose and the types of content used in ASAT rule documentation, we employ an informal open card-sorting~\cite{zimmermann_card-sorting_2016} methodology coupled with a saturation process similar to the one used in~\Cref{sec:analysis}. The objective is twofold:~identify \emph{how} the content is materialized in the documentation (\eg text, source code, images, \etc), and what is the \emph{purpose} of the content.

Traditional card sorting requires printing fragments of the documentation content on cards and physically regrouping them by common themes.
Unfortunately, the large amount of content present in our sample (more than a hundred rules) makes this methodology impractical.

To simplify the process, we go over rules incrementally in an arbitrary order, limiting the open card sorting process to the content of these rules relevant to the \textbf{Comprehension} theme. We end the process when no new theme emerges after five successive rules. 

Two of the authors, who participated also in the construction of the nomenclature and were already familiar with the content of the documentation, conducted the open card sorting during a collaborative session, in which they reviewed a few dozen rules.
At the end of the session, they obtained the following taxonomy:

\begin{itemize}
    \item \textbf{Purpose}
    \begin{itemize}
        \item \what: What triggers the activation of this rule and how to recognize violating code?
        \item \why: Why does this rule matter, and why should it be enforced?
        \item \fix: How should violating code be fixed to comply with this rule?
    \end{itemize}
    \item \textbf{Content type}
    \begin{itemize}
        \item \txt: Free-form prose text
        \item \source: Source code written in (one of) the programming language targeted by the ASAT, possibly including some prose embedded as comments
        \item \link: Hyperlinks to other documentation, web pages, PDFs, \etc
    \end{itemize}
\end{itemize}

\Cref{fig:rule_taxo_coded} shows an example rule from \texttt{Checkstyle} 
with its purposes and content types.
Its description employs a mix of \txt and \link to document the \what and \why purposes. The \source, on the other hand, documents the \what and \fix purposes via examples of compliant and non-compliant code.

\begin{table*}
    \centering
    \small
    \caption{Percentage of presence of each taxonomy purposes regarding the type of content for each ASAT}
    \begin{tabular}{@{}rlcccccccccc@{}}
        \toprule
            & & & \multicolumn{3}{c}{\textbf{Text}} & \multicolumn{3}{c}{\textbf{Code}} & \multicolumn{3}{c}{\textbf{Link}} \\
        \cmidrule(r){4-6} \cmidrule(lr){7-9} \cmidrule(l){10-12}
            \textbf{Language} & \textbf{ASAT} & \textbf{\# Rules} & \textit{What} (\%) & \textit{Why} (\%) & \textit{Fix} (\%) & \textit{What} (\%) & \textit{Why} (\%) & \textit{Fix} (\%) & \textit{What} (\%) & \textit{Why} (\%) & \textit{Fix} (\%) \\
        \midrule
            \multirow{2}{*}{C / C++}    & OCLint       & 6    & \tableCell{100} & \tableCell{33}  & \tableCell{0}   & \tableCell{100} & \tableCell{0}  & \tableCell{0}   & \tableCell{100} & \tableCell{0}  & \tableCell{0}  \\
                                        & Cppcheck     & 6    & \tableCell{100} & \tableCell{83}  & \tableCell{0}   & \tableCell{100} & \tableCell{0}  & \tableCell{17}  & \tableCell{83}  & \tableCell{83} & \tableCell{0}  \\
            \multirow{2}{*}{C\#}        & Gendarme     & 9    & \tableCell{100} & \tableCell{67}  & \tableCell{78}  & \tableCell{89}  & \tableCell{22} & \tableCell{89}  & \tableCell{0}   & \tableCell{0}  & \tableCell{0}  \\
                                        & Roslynator   & 10   & \tableCell{10}  & \tableCell{0}   & \tableCell{0}   & \tableCell{100} & \tableCell{0}  & \tableCell{90}  & \tableCell{10}  & \tableCell{0}  & \tableCell{0}  \\
            \multirow{2}{*}{Java}       & Checkstyle   & 6    & \tableCell{100} & \tableCell{50}  & \tableCell{50}  & \tableCell{100} & \tableCell{0}  & \tableCell{100} & \tableCell{100} & \tableCell{0}  & \tableCell{0}  \\
                                        & SpotBugs     & 6    & \tableCell{100} & \tableCell{83}  & \tableCell{100} & \tableCell{0}   & \tableCell{0}  & \tableCell{0}   & \tableCell{33}  & \tableCell{17} & \tableCell{0}  \\
            \multirow{2}{*}{JS / TS}    & ESLint       & 10   & \tableCell{100} & \tableCell{80}  & \tableCell{90}  & \tableCell{100} & \tableCell{10} & \tableCell{100} & \tableCell{100} & \tableCell{0}  & \tableCell{0}  \\
                                        & RSLint       & 8    & \tableCell{100} & \tableCell{100} & \tableCell{50}  & \tableCell{100} & \tableCell{36} & \tableCell{75}  & \tableCell{100} & \tableCell{0}  & \tableCell{0}  \\
            \multirow{2}{*}{PHP}        & PHP CS Fixer & 7    & \tableCell{100} & \tableCell{0}   & \tableCell{57}  & \tableCell{100} & \tableCell{0}  & \tableCell{100} & \tableCell{0}   & \tableCell{0}  & \tableCell{0}  \\
                                        & Psalm        & 7    & \tableCell{100} & \tableCell{14}  & \tableCell{14}  & \tableCell{100} & \tableCell{0}  & \tableCell{29}  & \tableCell{0}   & \tableCell{0}  & \tableCell{14} \\
            \multirow{2}{*}{Python}     & Pylint       & 10   & \tableCell{100} & \tableCell{10}  & \tableCell{40}  & \tableCell{80}  & \tableCell{0}  & \tableCell{80}  & \tableCell{100} & \tableCell{0}  & \tableCell{0}  \\
                                        & Flake8       & 6    & \tableCell{100} & \tableCell{17}  & \tableCell{83}  & \tableCell{100} & \tableCell{0}  & \tableCell{100} & \tableCell{83}  & \tableCell{0}  & \tableCell{0}  \\
            \multirow{2}{*}{Ruby}       & RuboCop      & 9    & \tableCell{100} & \tableCell{44}  & \tableCell{33}  & \tableCell{100} & \tableCell{0}  & \tableCell{89}  & \tableCell{44}  & \tableCell{0}  & \tableCell{0}  \\
                                        & Brakeman     & 7    & \tableCell{100} & \tableCell{86}  & \tableCell{57}  & \tableCell{71}  & \tableCell{0}  & \tableCell{14}  & \tableCell{71}  & \tableCell{57} & \tableCell{43} \\
            \multirow{2}{*}{Multi}      & Semgrep      & 6    & \tableCell{100} & \tableCell{67}  & \tableCell{67}  & \tableCell{100} & \tableCell{0}  & \tableCell{67}  & \tableCell{100} & \tableCell{50} & \tableCell{50} \\
                                        & SonarLint    & 6    & \tableCell{100} & \tableCell{100} & \tableCell{83}  & \tableCell{100} & \tableCell{0}  & \tableCell{100} & \tableCell{17}  & \tableCell{17} & \tableCell{17} \\
        \midrule
        Total & & 119 & \tableCell{93} & \tableCell{50} & \tableCell{50} & \tableCell{91} & \tableCell{5} & \tableCell{69} & \tableCell{58} & \tableCell{12} & \tableCell{7} \\
        \bottomrule
    \end{tabular}
    \label{tab:taxo_for_dataset}
\end{table*}

\subsection{Validation}
\label{sec:taxo-validation}

We validate the completeness and objectivity of our taxonomy by calculating the agreement of independently annotating a set of rules. Note that this is an internal validation of our taxonomy; our survey (\Cref{sec:survey}) validates it with external respondents.

The first author selects 12 rules from the considered ASATs, making sure to select non-trivial\footnote{An example of a trivial rule is \url{https://rslint.org/no-await-in-loop/}} rules covering the main categories of ASAT rules identified by Vassolo et~al.\xspace~\cite{vassallo_how_2020}: naming and style, correctness, performance, and security.
The first author then manually removes (when present) from their documentation the content related to \textbf{Usage} and \textbf{Metadata}---irrelevant to our taxonomy---keeping only the content related to \textbf{Comprehension}.
Finally, the first author annotates the rules by highlighting the elements that address the \what, \why, and \fix purposes, as well as the content types used, as shown in~\Cref{fig:rule_taxo_coded}.

The twelve rules are split randomly among three of the remaining authors, who follow the same rating process on four rules each, ensuring that each rule is annotated by two independent raters. We evaluate our taxonomy on these rules by assessing i)~to what extent the \what, \why, and \fix purposes are necessary and sufficient to rate all the documentation content (completeness), and ii)~to what extent independent raters reliably agree on the rating of documentation content (objectivity).

\paragraph{Completeness.}  All raters used all purposes on all the rules. All raters used at least one purpose on all \txt content, most \links, and the majority of \source snippet.
Some parts of the \source snippets were not rated as they do not relate to any purpose but rather serve as boilerplate code to ensure that the snippets are syntactically valid (for instance the class declaration in~\Cref{fig:rule_taxo_coded}).
One rater found that some \links were too general to be actionable and that it was not clear how they relate to the rule documentation (\eg a link to a list of the ten most critical vulnerabilities in web applications).\footnote{\url{https://owasp.org/www-project-top-ten/}}

\paragraph{Objectivity.} We measure the agreement among raters using Cohen's kappa for the textual content, at the word level with regard to the purposes. (the content type being completely objective).
Cohen's kappa is suitable for this situation as there are two raters per word, and each word is rated with a single purpose.
For the \source and \links, however, the same link or snippet may be associated with multiple labels.
We use the better-suited weighted Fleiss' kappa with a weight computed using the MASI distance between sets of labels~\cite{passonneau_measuring_2006}.
\begin{itemize}
    \item \txt: we obtain kappa values of $0.3$, $1$, and $0.741$ between the first author and the three other raters, indicating a rather strong agreement on textual content, except with one rater. The rater with the lowest agreement value performed its rating at the sentence level, while the other raters coded at the finer-grained level of individual words.
    \item \source: the kappa values are $1$, $0.71$, and $1$, indicating a very strong agreement on this type of content.
    \item \links: we obtain kappa values of $0.04$, $0.33$, and $0.14$, suggesting a low agreement between the raters for this type of content. The disagreements are due to: i)~some raters only rated using the context in which the link was used, without looking at its content, to perform the rating ii)~some raters did not assign any rating to very general links, while others assigned all the possible ratings.
\end{itemize} 
Since the agreement is overall high, with clear reasons for the disagreements we observe, our results indicate that our taxonomy is suitable for classifying ASAT documentation content. 

\subsection{Results}
\label{sec:taxo-results}

The first author applies the taxonomy to the 119 rules of \Cref{tab:rules_extracted} to highlight the content types and purposes for all documentation content pertaining to the \textbf{Comprehension} category.
With regard to the objectivity issues identified in~\Cref{sec:taxo-validation}, the first author applies a fine-grained strategy to rate the \txt content (rating at the word level), and an optimistic strategy to rate \links (assigning multiple purposes to general documents using their content). \Cref{tab:taxo_for_dataset} shows the percentage of rules, for each ASAT, that document the \what/\why/\fix purposes for \txt, \source, and \links.

\paragraph{Purposes.} Out of the 119 analyzed rules, 119 (100\%) document the \textit{What} purpose, 60 (50\%) document the \textit{Why} purpose, and 92 (77\%) document the \textit{Fix} purpose, regardless of the content type. Breaking down by content type, we see:
\begin{itemize}
    \item \what: 110 rules document it with \txt (93\%), 108 with \source (91\%), and 69 with \links (58\%).
    \item \why: 60 rules document it with \txt (50\% overall, 100\% when present), 6 with \source (5\% overall, 10\% when present), and 14 with \links (12\% overall, 23\% when present). 
    \item \fix:  59 rules document it with \txt (50\% overall, 64\% when present), 82 with \source (69\% overall, 90\% when present), and 8 with \links (7\% overall, 9\% when present). 
\end{itemize}


\paragraph{Content types.} \txt is present in 110 of 119 rules (92\%), \source  in 108 (91\%), and \links in 70 (59\%). Breaking down by purpose, we see:
\begin{itemize}
    \item \txt documents the \what purpose in 100\% of cases, the \why in 55\% of cases, and the \fix in 54\% of cases.
    \item \source documents the \what purpose in 100\% of cases, the \why in 6\% of cases, and the \fix purpose in 76\% of cases.
    \item \links document the \what purpose in 99\% of cases, the \why in 20\% of cases, and the \fix in 11\% of cases.
\end{itemize}

\paragraph{By tool.} As seen in the nomenclature, there is some variability. While All ASATs document the \what purpose, one ASAT does not document the \fix purpose (OCLint) and some rarely document it (Cppcheck, Psalm). Finally, some ASATs do not document the \why purpose (Roslynator, PHP CS Fixer) and some rarely document it (Psalm, Pylint, Flake8, OCLint).

\begin{finding}
\textbf{Finding \#2:} ASAT documentation has three main purposes: while the \what is systematically documented, the \fix is often documented (77\%), and the \why is documented only half of the time (50\%). The \what is documented with \txt (92\%), \source (91\%) or \links (58\%), the \why with \txt (100\%), \links (23\%) or \source (10\%), and the \fix with \source (90\%), \txt (69\%) or \links (9\%).

\end{finding}

\section{Questionnaire Survey}
\label{sec:survey}

In this section, we evaluate whether the documentation of ASAT rules meet the expectations of their users, as well as validating the taxonomy with said users.
Specifically, we assess whether the purposes documented in the rules and their incarnation as text, source code, and hyperlinks satisfy the developers facing them.
To answer these questions, we design an anonymous questionnaire survey with a mix of open-ended and closed-ended questions shared with industrial partners and fellow researchers.
\Cref{sec:survey-design} presents the design of our survey and \Cref{sec:survey-participants} gives an overview of the participants and the analysis methodology.
\Cref{sec:survey-results} details the quantitative results obtained for closed-ended questions and \Cref{sub:survey/qualitative} the qualitative results emerging from the open-ended questions.
The survey, responses, and plots we discuss in this section are available on an interactive companion webpage.\footnote{\url{https://icpc2024-asats.github.io?page=survey}}

\subsection{Survey Design}
\label{sec:survey-design}

The survey opens with a welcome message detailing its goals, authors, estimated completion time, and data policy.
The remainder of the survey revolves around four parts:~developer profile, taxonomy evaluation, rules analysis, and general feedback.
\Cref{tab:survey_questions} shows the questions in the survey.

\begin{table*}[tb]
    \small
    \centering
    \caption{Our questionnaire survey's questions. We used \emph{linter} instead of ASAT as it is more popular among developers.}
    \begin{tabular}{@{}lllc@{}}
                                             & \textbf{Question} & \textbf{Type} & \textbf{Mandatory} \\
        \midrule
            \multirow{5}{*}{\textit{Developer profile}}             & What is your experience as a developer? & Single choice & \ding{51} \\
                                                 & Which of the following languages do you use regularly? & Multiple choices & \ding{55} \\
                                                 & Do you know what a linter is? & Yes/No & \ding{51} \\
                                                 & Do you use a linter on some of your projects? & Yes/No & \ding{55} \\
                                                 & Which of the following linters were used in those projects? & Multiple choices & \ding{55} \\
        \midrule
            \multirow{9}{*}{\textit{Taxonomy evaluation}} & Rate the usefulness of each purpose in the documentation of a linter & Single choice for each purpose & \ding{51} \\
                                                 & For the \what purpose, why do you think it is (not) important to be & \multirow{2}{*}{Open-ended} & \multirow{2}{*}{\ding{55}} \\
                                                 & present in the documentation? & & \\
                                                 & For the \why purpose, why do you think it is (not) important to be & \multirow{2}{*}{Open-ended} & \multirow{2}{*}{\ding{55}} \\
                                                 & present in the documentation? & & \\
                                                 & For the \fix purpose, why do you think it is (not) important to be & \multirow{2}{*}{Open-ended} & \multirow{2}{*}{\ding{55}} \\
                                                 & present in the documentation? & & \\
                                                 & Do you think that there are other purposes that a linter documentation & \multirow{2}{*}{Open-ended} & \multirow{2}{*}{\ding{55}} \\
                                                 & should have? & & \\
        \midrule
            \multirow{9}{*}{\textit{Rules analysis}}       & Have you ever seen this rule? & Yes/No & \ding{51} \\
                                                 & For the rule and taxonomy provided, evaluate for each type of content & \multirow{2}{*}{Single choice for each type} & \multirow{2}{*}{\ding{51}} \\
                                                 & its importance to explain the What purpose & & \\
                                                 & For the rule and taxonomy provided, evaluate for each type of content & \multirow{2}{*}{Single choice for each type} & \multirow{2}{*}{\ding{51}} \\
                                                 & its importance to explain the Why purpose & & \\
                                                 & For the rule and taxonomy provided, evaluate for each type of content & \multirow{2}{*}{Single choice for each type} & \multirow{2}{*}{\ding{51}} \\
                                                 & its importance to explain the Fix purpose & & \\
                                                 & For the rule and taxonomy provided, indicate your satisfaction level & \multirow{2}{*}{Single choice for each purpose} & \multirow{2}{*}{\ding{51}} \\
                                                 & on the quality of the documentation for each purpose & & \\
        \midrule
            \multirow{2}{*}{\textit{General feedback}}    & Please comment freely on the linters documentation you saw: what you & \multirow{2}{*}{Open} & \multirow{2}{*}{\ding{55}} \\
            & appreciated, disliked, and how it compared with your expectations. & & \\
        \bottomrule
    \end{tabular}
    \label{tab:survey_questions}
\end{table*}

\paragraph{Developer profile}
To establish the profile of our participants, we ask about their experience as developers and which programming languages they use regularly.
Participants may pick between four groups:~\textit{Novice} (0 to 4 years of experience), \textit{Junior} (5--9), \textit{Confirmed} (10--19), and \textit{Senior} (20+).
They select or enter their preferred programming languages from an open-ended list.
As our study focuses explicitly on ASATs, we ask the participants about their experience with linters.\footnote{The survey uses the term linter rather than ASAT, as it is much more popular.}
The first question asks whether they know what a linter is, the second whether they use linters on some of their projects, and the last one which linters they use (picked from the 16 ASATs we study in this paper or entered manually).

\paragraph{Taxonomy evaluation}
In this part, participants are shown a screenshot of the \texttt{FetchEnvVar} rule from RuboCop\footnote{\url{https://docs.rubocop.org/rubocop/cops_style.html\#stylefetchenvvar}} which serves as an illustration to the terminology we use in the survey (linter, rule, compliant code, non-compliant code).
Then, we introduce the participants to the terms of our taxonomy and their definition:~purposes (\what/\why/\fix) and content types (\txt, \source, \links).
We then ask the participants to evaluate the importance of each purpose in the documentation of ASAT rules.
We employ an asymmetric survey response scale inspired by Kano~\etal~\cite{n_attractive_1984} and adapted by Begel~\etal~for software engineering~\cite{begel_analyze_2014}:~\textit{Essential}, \textit{Worthwhile}, \textit{Unimportant}, \textit{Unwise}, \textit{I don't understand}.
For each purpose, we include an additional open-ended question asking why its presence in the documentation is or is not important.
The final open-ended question asks the participants whether ASATs should document additional purposes, which we may have missed in our taxonomy.

\begin{figure}
\includegraphics[width=\linewidth]{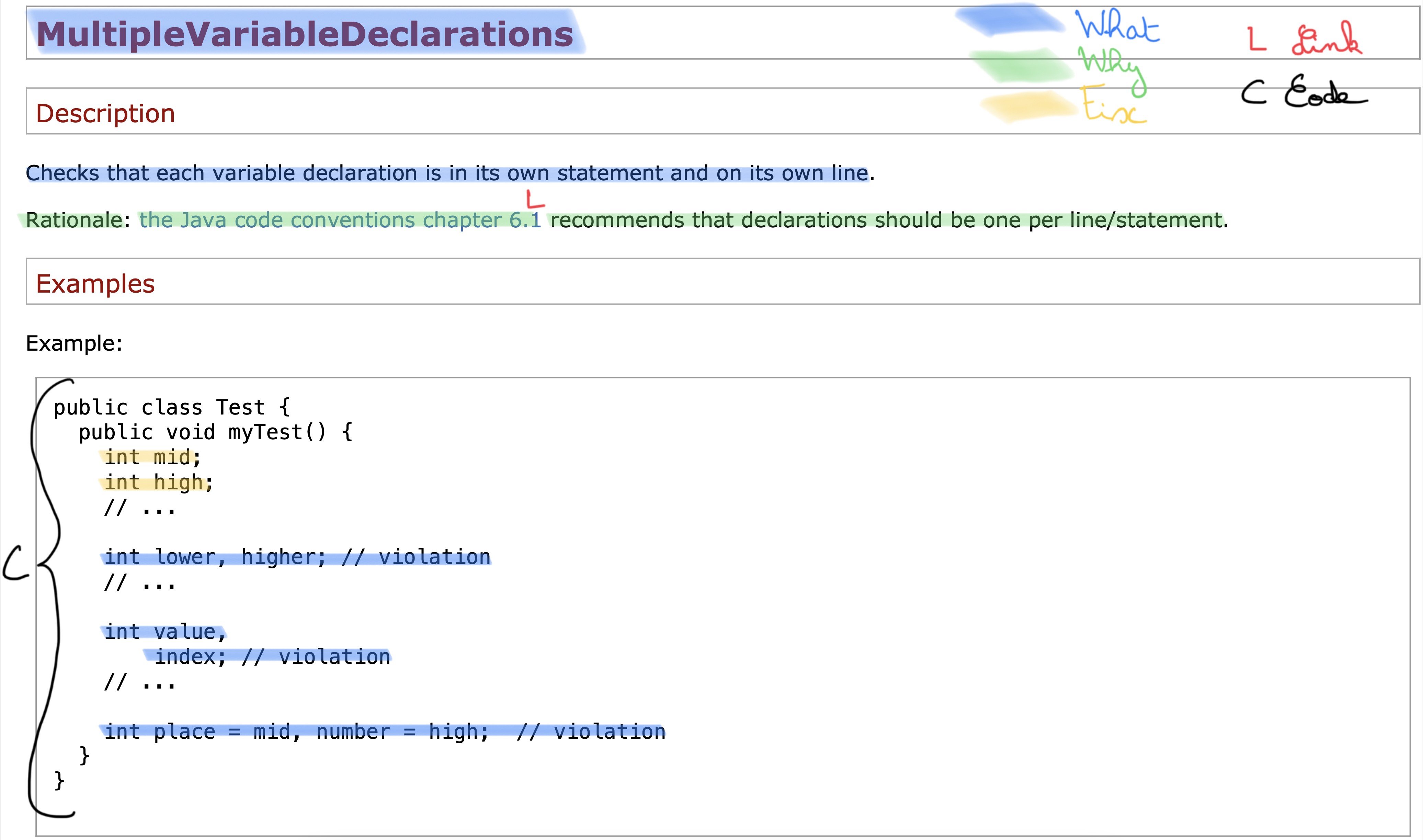}
  \caption{Taxonomy applied to rule \textit{MultipleVariableDeclarations} from \texttt{Checkstyle}}
\label{fig:rule_taxo_coded}
\end{figure}

\paragraph{Rules analysis}
In this part, participants are tasked to evaluate the documentation of concrete ASAT rules.
For each of the 12~rules used to validate our taxonomy in \Cref{sec:taxo-validation}, we create a bespoke page in the survey.
The page includes a screenshot of the rule's documentation, a link to its official documentation, and a series of questions regarding its quality.
As our goal is not to evaluate the ability of participants to annotate the rule with our taxonomy, the screenshot includes information regarding its purposes and content types (as shown in \Cref{fig:rule_taxo_coded}), agreed upon by four authors (the screenshots annotated are available in our replication kit).\footnote{\url{https://icpc2024-asats.github.io/file/survey_rules_annotated.pdf}}

The first question asks whether the participant already knows the rule.
Then, for each purpose and each content type, a question asks to evaluate the importance of this content type to document this specific purpose using the asymmetric scale introduced above.
Finally, the last series of questions asks the participant to judge the overall quality of the documentation for each purpose, using a symmetric scale to measure satisfaction:~\textit{Very satisfied}, \textit{Satisfied}, \textit{Neither satisfied nor dissatisfied}, \textit{Dissatisfied}, \textit{Very dissatisfied}.
Participants may answer \textit{Not present} to any question, indicating that there is no content of the appropriate type or that the rule does not document the purpose of interest.
The \textit{Not present} answer also serves as a quality check, as shown later in \Cref{sec:survey-participants}.

When participants complete the first two parts of the survey, one rule out of the 12 is drawn randomly and displayed.
They can then opt-in to analyze another rule, drawn randomly from the remaining ones until there is no more rule to examine.
When a participant evaluates all the rules or opts out, he is redirected to the last part of the survey.

\paragraph{General feedback}
This final part consists of a single open-ended question asking the participants to comment on the documentation of the rules they evaluated.
This question is designed to put our taxonomy aside and invite the participants to share their opinions more freely:~what they liked and disliked about the rules and their documentation, and how it compared with their expectations.

Finally, we proceeded to a pilot testing of the survey with three software engineering researchers external to the study, who provided feedback about the phrasing of the questions and a time estimation to complete it.

\subsection{Participants and Methodology}
\label{sec:survey-participants}

We primarily shared the survey with industrial partners and fellow researchers through direct contact and mailing lists.
We also publicized it on social and professional networks (Twitter, LinkedIn, Slack).
The survey was available online on a self-hosted LimeSurvey instance from July 4 to October 19, 2023.
Overall, we received a total of 179 anonymous answers.
The participants left at different stages:~179 entered their developer profile, 119 evaluated our taxonomy, 85 evaluated at least one rule (for a total of 289 rule evaluations), and 26 answered the last open-ended question.

Our methodology to sanitize the data and analyze the responses is as follows.
First, we clean up the responses and attempt to remove noise.
As mentioned earlier, the participants can answer \textit{Not present} when a type of content or a given purpose is missing from the documentation of the rule they are evaluating.
We observe that, in some cases, participants marked some type of content or purpose as \textit{Not present} although it was present and marked as such in the screenshot.
For instance, some participants answered that the \fix purpose for the \source was not present in the screenshot of \Cref{fig:rule_taxo_coded}.
In this case, as a sanity measure, we discard the participant's answers related to this purpose for the given rule, for all content types.
We apply the same filter when a participant provides an evaluation for a purpose that is not present in the rule presented to them.
We obtained 225 evaluations for the \what purpose, 91 for the \why purpose, and 161 for the \fix purpose.

Second, we use thematic analysis~\cite{guest_applied_2023} to extract codes from the answers to open-ended questions.
Two authors read these answers and assigned codes.
Then, the four first authors gather to harmonize the codes under higher-level themes, discussed in \Cref{sub:survey/qualitative}.
Overall, we obtained 33 answers regarding whether ASAT rules should document other purposes, 56 responses evaluating the importance of each purpose in the documentation (168 in total), and 26 responses to the \textit{General feedback} question.

\subsection{Quantitative Analysis}
\label{sec:survey-results}

In this section, we review the responses to closed-ended questions.
We only include the responses of the 85 participants that have evaluated at least one rule.

\paragraph{Developer profile}
The distribution of participants in terms of experience is fairly even.
Of the 85 participants, 37 have less than 5 years of experience as a developer (novices), 22 have between 5 and 9 years (juniors), and 26 have more than 10 years (seniors). The most used programming language is C/C++ (55\%), closely followed by Python (54\%), JavaScript and TypeScript (44\%), and Java (42\%).
The remaining languages are used by less than 15\% of respondents.

A large majority (81\%) of participants do know what an ASAT is; the proportion grows with experience (73\% of novices, 82\% of juniors, and 92\% of seniors).
The same trend is found for ASAT usage: 65\% of the participants use or used ASATs in their projects (46\% for novices, 68\% for juniors, and 88\% for seniors).
There is also a noticeable imbalance in ASAT use depending on the programming languages they use: 53\% of C/C++ developers use an ASAT, 61\% of Python developers, 64\% of Java developers, and 81\% of JavaScript and TypeScript developers.
A plausible explanation is that ESLint is often bundled by default when initializing JavaScript and TypeScript projects. 
When participants use ASATs in their projects, the most popular one is indeed ESLint (41\%), followed by Pylint (21\%), and SonarLint (19\%). The remaining tools are used by less than 10\% of participants; three respondents mention additional ASATs they use---clang tidy, Fortify and OCaml platform---indicating that our selection of ASATs reflects the ones used in practice. Overall ASAT usage reflects language use, with the exception of C/C++: since developers use more than one language, C/C++ users tend to use ASATs with other programming languages. 

\begin{figure}[tb]
    \centering
    \begin{subfigure}{\linewidth}
        \includegraphics[width=\linewidth]{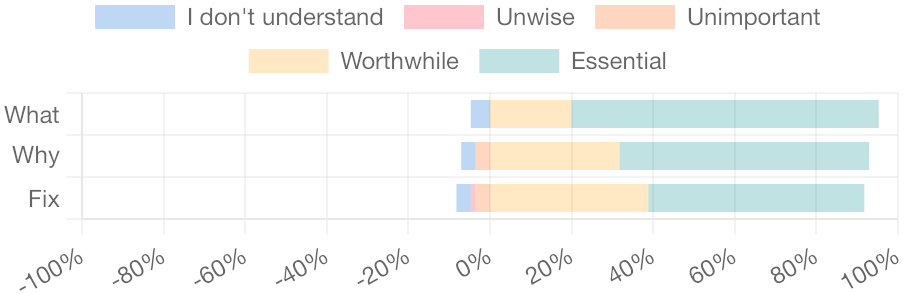}
        \caption{Usefulness of each purpose in the documentation}
        \label{fig:purpose_usefulness}
    \end{subfigure}
    \begin{subfigure}{\linewidth}
        \includegraphics[width=\linewidth]{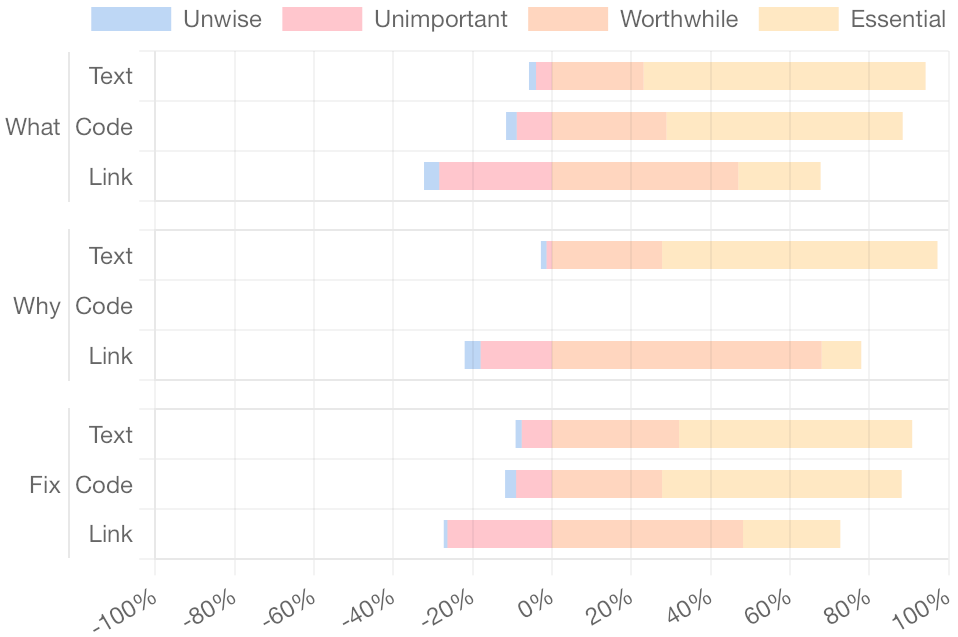}
        \caption{Importance of each type of content to document each purpose}
        \label{fig:purpose_importance}
    \end{subfigure}
    \begin{subfigure}{\linewidth}
        \includegraphics[width=\linewidth]{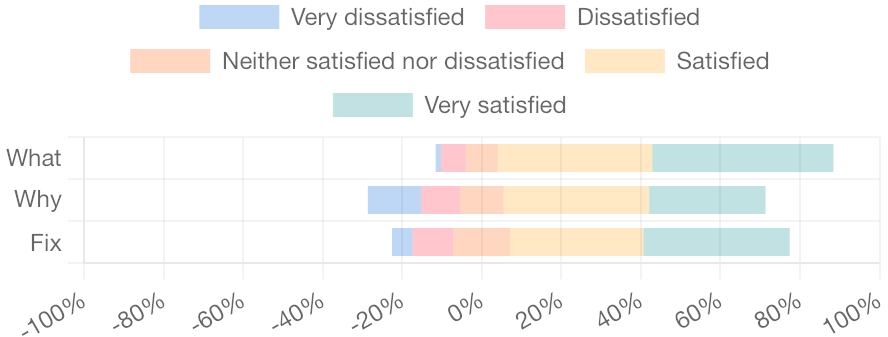}
        \caption{Quality of the documentation for each purpose}
        \label{fig:purpose_satisfaction}
    \end{subfigure}
    \caption{Participants' evaluation of the usefulness, importance, and quality of the content types and purposes}
\end{figure}

\paragraph{Taxonomy evaluation}
In this part, we ask participants to judge the usefulness of the \what, \why, and \fix purposes in the documentation of ASATs, independent of any particular rule.
\Cref{fig:purpose_usefulness} shows the results: the participants strongly expect each of the purposes to be present and documented.
While the \what and \why purposes are largely judged as essential, the \fix purpose appears slightly less essential to participants, but still worthwhile.
In particular, we note the importance of the \why purpose to participants, which indicates that rule documentation should not only document the problem and its solution but also the rationale motivating the rule; in contrast, \emph{only half of the rules} we analyzed had a rationale documented (\Cref{tab:taxo_for_dataset}).
When grouping answers by developer experience, programming language, or other profile criteria we do not observe any major differences.

\paragraph{Rules analysis}
The 85 participants analyzed a total of 289 rules, with a mean of $3.4$ rules per participant.
Each of the 12 rules has been evaluated by 19 to 29 different participants.
\Cref{fig:purpose_importance} shows how the participants rate the importance of each type of content to document each purpose in the rules they examined.
If a rule does not include a particular type of content to document a given purpose, and the participant marks it as \textit{Not present}, we omit this data point as it does not convey any positive or negative judgment.
This explains why there is no evaluation of \source to document the \why purpose, as none of the 12 rules document it with code.

Participants evaluate the importance of the \what purpose positively (\textit{Essential} or \textit{Worthwhile}), regardless of its incarnation (94\% for \txt, 88\% for \source, and 68\% for \links).
While text and source code are seen as essential, the \links are mainly judged as worthwhile.
For the \why purpose, participants mostly evaluate \txt as essential (69\%) and \links as worthwhile (68\%).
For the \fix purpose, participants mostly evaluate \txt (59\%) and \source (60\%) as essential, and \links as worthwhile (48\%).

Participants evaluate \txt and \source very positively to explain the three purposes.
This suggests that a combination of text and source code might be the best choice to document ASAT rules.
Participants also evaluate \links positively, with a minimum of 68\% of positive evaluations for each purpose.
Yet, most participants evaluate them as \textit{Worthwhile} rather than \textit{Essential}. Moreover, a significant portion judges them as \textit{Unimportant}: three times or more than \txt or \source---and more than they are judged \textit{Essential}.
A possible reason is that visiting external resources disrupts the reading flow and that important information may be lost among other resources, particularly if linking to larger documents.
A solution could be to extract and distill the important information from external resources into the documentation, and cite it as a source.

\begin{finding}
\textbf{Finding \#3:} Text and source code are best suited to document the \what and \fix purposes, and good documentation for an ASAT rule should include both.
\txt is the best-suited medium to document the \why purpose. \links are rarely seen as essential and should be used sparingly.
\end{finding}

Finally,~\Cref{fig:purpose_satisfaction} displays the satisfaction of participants regarding the quality of the documentation in the rules they evaluated, for each purpose.
We observe that the participants are mostly satisfied with the quality of the documentation for the \what (84\%) and \fix purposes (70\%).
Yet, we observe that almost a quarter of the participants are not satisfied with the quality of the documentation for the \why purpose.
Worse, in 13\% of cases, participants are very dissatisfied with its quality. This strongly contrasts with the usefulness evaluation emitted by participants in \Cref{fig:purpose_usefulness}. 

\begin{finding}
\textbf{Finding \#4:} Participants express a strong interest in understanding the rationale behind ASAT rules when reading their documentation (\why).
Yet, \textbf{Finding \#1} indicates that only 50\% of ASAT rules document the \why purpose; when present, participants are dissatisfied with how it is documented in close to 25\% of the cases. Clearly, \emph{documentations should consistently include and explain the \why aspect}.
\end{finding}

\subsection{Qualitative Analysis}
\label{sub:survey/qualitative}

\newcommand{\code}[2]{\textbf{#1 (#2)}}
\newcommand{\refcode}[1]{\emph{#1}}
\newcommand{\qlquote}[1]{\emph{``#1''}}

In the following, we summarize free-form comments based on our coding. Codes mentioned for the first time include their frequency \code{like this}{0}. Code referenced after being introduced once is \refcode{like this}. We do not report on ``obvious'' codes (e.g., \code{understanding the rationale}{29} for the why purpose). 

\paragraph{Transversal theme: learning}
One of the most salient themes, across all three purposes (\code{learning-what}{11}, \code{learning-fix}{13}, and particularly \code{learning-why}{21}), is \emph{learning}: 45 comments touched on that theme in one way or another. This is sometimes phrased as self-improvement of the developer's skills, particularly for beginners or (more occasionally) for onboarding new team members. Comments on the \what aspect mention the need to understand the error to not repeat it (thus improving skills), as well as team aspects (\qlquote{because it facilitates the integration of new members}). Comments on the \why are the most prevalent. Explaining the rationale for an error is important to understand its purpose and importance, and from that, remembering it: \qlquote{If I don't know why, then I don't know why it's a bad thing and I cannot improve as a developer}. Finally, the \fix is more immediate. Once the problem is known and understood, learning of possible solutions is valuable: \qlquote{It is likely the coder introduced a bad pattern/error due to lack of expertise; as such, it would be wrong to assume he/she will know how to fix it}.

\paragraph{Transversal theme: saving time} 
Respondents emphasized efficiency in all three aspects (\code{saving time-what}{2}, \code{saving time-why}{3}, particularly \code{saving time-fix}{13}, \code{automated fixes}{4}). For instance, one respondent wants to \qlquote{understand in seconds what a lint is about}, which requires clear and concise explanations. Missing information in the \why prevents one from deciding whether to act on a warning (\qlquote{I will probably lose time looking it up on the internet. Also I will be less motivated to fix it}). Since fixes are the most actionable, there is more demand to have standard solutions available to solve the issue quickly, rather than searching for information on the web or asking teammates. Having to search for the information is perceived to increase the chance a warning is ignored. Going further, the logical end is automation: \qlquote{ideal thing is to just click a button to 'autofix' the issue, when available}.


\paragraph{Aspects specific to the \what.} 
The \what provides understanding of what triggers a rule. It should be written clearly and concisely to this as easy as possible (thus saving time). The \what helps finding \code{where the warning is}{8}, which is not always obvious as \qlquote{many different things may be discussed/considered on a single piece of code}. A key step in finding where the error is, is to \code{relate the error to one's code}{12}: rules are either described in the abstract, or, at best, with an \code{example}{3}. Examples are preferred for this: \qlquote{simpler the exemple [sic], the easier it is to relate to one's code}. Another theme relates to \code{false positives and negatives}{5}. Rules are implemented by heuristics that may be imperfect especially for complex cases (e.g. regular expressions). Describing what triggers a rule in details is useful to disambiguate between true positives and false positives, as well as knowing cases that the rule can miss (false negatives).

\paragraph{Aspects specific to the \why.}
The \why is key to deciding whether to act on the warning, or not. Developers \code{analyse the tradeoffs and risks involved}{11}: the \why should in particular explain the severity of the warning: \qlquote{I can judge whether the criticality justify modifying this piece of code. I may choose to disregard the rule if I judge it not worthwhile}. In some cases, the rule's \code{relevance}{9} will be questioned (such as when it is a \refcode{false positives and negatives}, or when it is a matter of personal or team \code{preferences}{5}, rather than a real issue. Thus the \why should \code{motivate and justify the effort}{10} that will be invested in fixing the warning; needless to say, if said effort is low (\refcode{saving time} via good \refcode{examples} or \refcode{automated fixes}), it will be easier to act on it.

\paragraph{Aspects related to the \fix.} There is more debate as to the importance of the \fix, compared to the \what and the \why. Some respondents state \code{fixes are less or not important}{7} for several reasons: either because the rules are simple, the fixes would be too basic, or because the \why and \what are sufficient (\qlquote{in most cases previous information should be enough to infer how to fix}. For other respondents, \code{fixes are essential or very important}{10}. For them, documentation without a fix is not actionable: \qlquote{If the Fix is not here, understanding the what and the why lead us to nowhere}. Fixes \code{increase ASAT friendliness}{5}, and help \refcode{saving time}. Fixes are particularly \code{useful when they are not obvious}{4}, either due to lack of knowledge (\refcode{learning}) or due to their difficulty. Providing \code{examples}{10} is useful to clarify complex rules. Some respondents mention that \code{fixes may not be optimal}{7} given the context, at worse they can \qlquote{lead beginners to apply a cascade of bad decisions made to satisfy the linter}.

\paragraph{Other purposes for documentation.} Most respondents mentioned \code{no additional purpose}{14} than the \what, \why, and \fix. A few respondents did mention other documentation purposes, such as \code{exceptions, alternatives, and limitations}{3}, \code{risks}{3}, or \code{configuration}{2} of the ASAT warnings. Indeed, we encountered these elements and included them in our nomenclature, but decided to exclude them from the remainder of our analysis as they dealt with more operational aspects.


\paragraph{Additional comments on the documentation.} Several participants emphasized in their free-form comments some topics that emerged earlier. Several respondents highlighted the \code{need for a summary}{7} (\qlquote{Sometimes too much text which does not encourage taking the time to read and therefore deal with the error} or a \code{structured template}{6} where \qlquote{the what, the why and the fix are clearly separated}. Others reiterated the need to \code{use code examples}{4}, including adding examples of both compliant and non-compliant code, or to \code{avoid links}{2} (\qlquote{external links are almost never useful}). 

\begin{finding}
    \textbf{Finding \#5:} ASAT documentation has important learning purposes (particularly the \why). Respondents read the \what to understand the error, the \why to decide whether to address the warning, and the \fix to remove it. Missing elements make fixes less likely.
    Some respondents even suggested the use of a structured template to enforce the presence of every important purpose.
    Additionally, some respondents expressed concern about efficiency.
    A summary was suggested by some respondents to speed up the understanding of the rule.
    Respondents plebiscited the use of code examples and recommended to use external links sparingly to avoid breaking the reading flow.
\end{finding}

\section{Threats to Validity}
\label{sec:threats}

\paragraph{External validity}

Our study bears several threats with regard to external validity.
The first threat is that our corpus of ASATs and ASATs rules documentation is not representative, and is even biased toward popular ASATs.
Therefore, we have no guarantee that our results would generalize to the actual population, especially the percentages displayed in~\Cref{tab:rules_extracted} and~\Cref{tab:taxo_for_dataset}.
Another threat is that the respondents of our survey are not a random sample of the population of ASATs users. As a consequence, the results we obtained with our participants might not generalize, especially the percentages displayed in~\Cref{fig:purpose_usefulness,fig:purpose_importance,fig:purpose_satisfaction}.

\paragraph{Internal validity}

We extensively use qualitative methods to build our nomenclature and taxonomy (open card sorting~\cite{zimmermann_card-sorting_2016}) and analyze the open answers to the survey (thematic analysis~\cite{guest_applied_2023}).
It is well known that these methods can be affected by subjectivity~\cite{dowling_power_2005}.
As a consequence, different researchers might have obtained a different nomenclature, taxonomy, and other themes from the survey answers.
As a mitigation measure for the taxonomy, we performed an internal validation and a reality check in the survey where we observed that it was well understood by the respondents.
For the other results obtained from qualitative analysis, we systematically used harmonization sessions to reduce the subjectivity of our findings.

With regard to the answers of the participants to the survey, there is a chance that they did not fully understand the questions we asked.
This threat affects mostly the closed-ended questions where we cannot do any sanity check by looking at the answer.
The most difficult questions concerned the analysis of the usefulness of each type of content to document each purpose, as the participants needed to carefully analyze how we rated the content of the rules.
As a sanity check, we included a specific scale item (Not present) to double-check that the participants did understand our rating.
We used this sanity check to filter out incoherent answers.
However, this sanity check is not perfect and it is possible that participants answered these questions without understanding our rating.

\section{Related Work}
\label{sec:rw}

To the best of our knowledge, there is no prior work having studied the rules documentation of ASATs.
In the remainder of the section, we discuss the related work according to two topics: studies of ASATs and studies of software documentation.

\subsection{ASATs Studies}

Researchers have emphasized various benefits associated with the use of ASATs.
Tómasdóttir~\etal highlight eight such advantages, such as error prevention, keeping the code simple and consistent, or improving the efficiency of code review and discussions~\cite{tomasdottir_why_2017}.
ASATs can also automate compliance to a standard coding style~\cite{ayewah_using_2008}.


Several studies explore the reasons inhibiting the use of ASATs by developers.
Tómasdóttir~\etal highlight that the agreement on rules to enable, especially for pre-existing projects, is an important issue and makes it difficult for developers to follow the rules when they disagree~\cite{tomasdottir_why_2017}.
Other challenges were also raised: dealing with false positives~\cite{heckman_systematic_2011}, having better integration into the development process~\cite{johnson_why_2013} and missing quick fixes when rules are detected~\cite{nachtigall_large-scale_2022}.

The study from Novak \etal~\cite{novak_taxonomy_2010} produces a taxonomy of ASATs across 10 categories (such as rule domains or configurability).
However, ASAT documentation has not been investigated in their study.

The studies on the warnings (also called notifications) content of ASATs are closely related to ours, as warnings are the first piece of documentation that developers see when using an ASAT.
An important result is that their content might not be sufficiently helpful to developers~\cite{tahaei_security_2021,johnson_cross-tool_2016,nachtigall_large-scale_2022}.
Another finding is the need to have clear and concise warnings, especially when developers use ASATs in CLI~\cite{gorski_listen_2020}.
It aligns with our findings for ASAT rule documentation: the main information should be quickly available and understandable.
Buckers~\etal present a tool and its evaluation to help developers treat more efficiently the numerous notifications and their contents~\cite{buckers_uav_2017}.
Overall, our results are consistent with those of previous related studies on warnings~\cite{tahaei_security_2021, johnson_cross-tool_2016, do_why_2022, johnson_why_2013, nachtigall_large-scale_2022}.

Tahaei~\etal find that examples are more valued and links less valued, which echoes our study~\cite{tahaei_security_2021}. They focus on four security warnings, while we cover more domains beyond security and explore more rules and ASAT tools.

Johnson~\etal study notifications for an ASAT, a coverage tool, and a compiler (five examples each)~\cite{johnson_cross-tool_2016}. Since they focus on multiple tool types, ASAT-specific aspects (\eg examples) are less prominent. However, they identify ``Problem Importance Gaps'' and ``Problem Resolution Gaps'' that echo our \why and \fix purposes.

Nachtigall~\etal define the objective of warning messages as such: ``Warning messages [...] have to direct the developer’s attention to the detected issue and give information on what might be wrong, why it should be fixed, and how it could be fixed'', echoing our taxonomy. However, it is not clear how they came up with this definition, while we obtain it from an analysis of existing ASAT documentation and confront it to the expectations of developers.~\cite{nachtigall_large-scale_2022}

Do~\etal extract three features (among many others) expected by ASAT users from a tool and a literature review (called F3, F4, and F7 in their article), which are similar to explaining the \what, \why, and \fix purposes.
These three features are among the most important ones according to the developers they surveyed~\cite{do_why_2022}.
The other features are more tool-oriented since their study focuses on tools and not documentation.

Finally, Johnson~\etal make the following observation: ``Nineteen of our 20 participants, felt that many static analysis tools do not present their results in a way that gives enough information for them to assess what the problem is, why it is a problem, and what they should be doing differently'' which also echoes our taxonomy~\cite{johnson_why_2013}.

One important novel aspect of our study is that it is the only one that focuses on reference documentation, not warnings or tools.
It confirms that the findings of the study about warnings are transferable to reference documentation.
In addition, our results provide more evidence about what is present/missing from existing documentation both for the nomenclature and taxonomy, while the qualitative part dives deeper into these topics.

\subsection{Software Documentation}

Another related topic is the study of documentation of Application Programming Interfaces (APIs).
API documentation shares a common goal with ASAT documentation as they are meant to teach developers how to correctly use an API as quickly as possible~\cite {watson_api_2013}.
On the other hand, they focus on a single software component while ASAT documentation describes rules that apply in a wide range of contexts.
We also find similar expected criteria with the writing of the rules documentation, as the need to include short code snippets demonstrating API usage in context or code examples illustrating the best practices~\cite{monperrus_what_2012,cummaudo_what_2019}.

Aghajani~\etal performed an empirical study on software documentation resulting in a taxonomy of documentation issues~\cite{aghajani_software_2019}.
An interesting outcome is the identification of themes relatively close to ours, especially \emph{Information Content} similar to our \emph{Comprehension}.
They pursue their work in another article~\cite{aghajani_software_2020} by realizing two surveys, first to evaluate the relevance of issues they had found, then to explore the needs on type of documentation.
It results in guidelines to improve the state of the art around documentation.

Regarding software documentation in general, one major flaw in software documentation is the difficulty of keeping it updated through its lifetime~\cite{forward_relevance_2002}.
Some developers simply assume it is outdated and do not rely on it~\cite{roehm2012professional}.
At first glance, we did not see this problem in the documentation of ASAT rules maybe because they evolve less often, but it would be an interesting issue to study.
\section{Conclusion}
\label{sec:conclusion}

In this article, we explored how ASATs rules are documented, and contrasted them with developers' expectations through several contributions. We first studied how more than 100 rules---spanning 16 ASATs in multiple programming languages---are documented, leading to a nomenclature of documentation elements refined in a taxonomy of documentation purposes and content types.
We then use this taxonomy to contrast the documentation of 12 real-world rules with developer expectations via a survey involving 85 respondents who evaluated the rules 289 times.
Among other findings, we highlight issues with the \why purpose: despite being considered essential by developers to decide whether to act on a warning, half of the rules miss it; of the rest, a quarter of survey responses point at quality issues. We also find out that code examples in addition to text are attractive for documenting the \what and \fix purposes. Finally, some developers expressed concern about saving time, leading to the recommendation to include summaries, and reducing the use of external links that disrupt the reading flow.
In future work, we plan to extend our study to more ASATs and to more quality issues inside the documentation, such as outdatedness, as has been done for API documentation~\cite{aghajani_software_2019,aghajani_software_2020}.

\begin{acks}
We would like to thank Matias Martinez, Lina Ochoa Venegas, and Théo Zimmermann for their initial feedback on the online survey.
\end{acks}

\balance

\bibliographystyle{IEEEtran}
\bibliography{IEEEabrv,bibfile}

\begin{thebibliography}{10}
\providecommand{\url}[1]{#1}
\csname url@samestyle\endcsname
\providecommand{\newblock}{\relax}
\providecommand{\bibinfo}[2]{#2}
\providecommand{\BIBentrySTDinterwordspacing}{\spaceskip=0pt\relax}
\providecommand{\BIBentryALTinterwordstretchfactor}{4}
\providecommand{\BIBentryALTinterwordspacing}{\spaceskip=\fontdimen2\font plus
\BIBentryALTinterwordstretchfactor\fontdimen3\font minus
  \fontdimen4\font\relax}
\providecommand{\BIBforeignlanguage}[2]{{%
\expandafter\ifx\csname l@#1\endcsname\relax
\typeout{** WARNING: IEEEtran.bst: No hyphenation pattern has been}%
\typeout{** loaded for the language `#1'. Using the pattern for}%
\typeout{** the default language instead.}%
\else
\language=\csname l@#1\endcsname
\fi
#2}}
\providecommand{\BIBdecl}{\relax}
\BIBdecl

\bibitem{novak_taxonomy_2010}
J.~Novak, A.~Krajnc, and R.~Žontar, ``Taxonomy of static code analysis
  tools,'' in \emph{The 33rd {International} {Convention} {MIPRO}}, May 2010,
  pp. 418--422.

\bibitem{tomasdottir_adoption_2020}
K.~F. Tómasdóttir, M.~Aniche, and A.~Van~Deursen, ``The {Adoption} of
  {JavaScript} {Linters} in {Practice}: {A} {Case} {Study} on {ESLint},''
  \emph{IEEE Transactions on Software Engineering}, vol.~46, no.~8, pp.
  863--891, Aug. 2020, conference Name: IEEE Transactions on Software
  Engineering.

\bibitem{beller_analyzing_2016}
\BIBentryALTinterwordspacing
M.~Beller, R.~Bholanath, S.~McIntosh, and A.~Zaidman,
  ``\BIBforeignlanguage{en}{Analyzing the {State} of {Static} {Analysis}: {A}
  {Large}-{Scale} {Evaluation} in {Open} {Source} {Software}},'' in
  \emph{\BIBforeignlanguage{en}{2016 {IEEE} 23rd {International} {Conference}
  on {Software} {Analysis}, {Evolution}, and {Reengineering} ({SANER})}}.\hskip
  1em plus 0.5em minus 0.4em\relax Suita: IEEE, Mar. 2016, pp. 470--481.
  [Online]. Available: \url{http://ieeexplore.ieee.org/document/7476667/}
\BIBentrySTDinterwordspacing

\bibitem{mclean_comparing_2012}
R.~K. McLean, ``Comparing {Static} {Security} {Analysis} {Tools} {Using} {Open}
  {Source} {Software},'' in \emph{2012 {IEEE} {Sixth} {International}
  {Conference} on {Software} {Security} and {Reliability} {Companion}}, Jun.
  2012, pp. 68--74.

\bibitem{ashfaq_comparative_2019}
Q.~Ashfaq, R.~Khan, and S.~Farooq, ``A {Comparative} {Analysis} of {Static}
  {Code} {Analysis} {Tools} that check {Java} {Code} {Adherence} to {Java}
  {Coding} {Standards},'' in \emph{2019 2nd {International} {Conference} on
  {Communication}, {Computing} and {Digital} systems ({C}-{CODE})}, Mar. 2019,
  pp. 98--103.

\bibitem{habchi_adopting_2018}
\BIBentryALTinterwordspacing
S.~Habchi, X.~Blanc, and R.~Rouvoy, ``On adopting linters to deal with
  performance concerns in {Android} apps,'' in \emph{Proceedings of the 33rd
  {ACM}/{IEEE} {International} {Conference} on {Automated} {Software}
  {Engineering}}, ser. {ASE} '18.\hskip 1em plus 0.5em minus 0.4em\relax New
  York, NY, USA: Association for Computing Machinery, Sep. 2018, pp. 6--16.
  [Online]. Available: \url{https://doi.org/10.1145/3238147.3238197}
\BIBentrySTDinterwordspacing

\bibitem{tahaei_security_2021}
\BIBentryALTinterwordspacing
M.~Tahaei, K.~Vaniea, K.~K. Beznosov, and M.~K. Wolters,
  ``\BIBforeignlanguage{en}{Security {Notifications} in {Static} {Analysis}
  {Tools}: {Developers}’ {Attitudes}, {Comprehension}, and {Ability} to {Act}
  on {Them}},'' in \emph{\BIBforeignlanguage{en}{Proceedings of the 2021 {CHI}
  {Conference} on {Human} {Factors} in {Computing} {Systems}}}.\hskip 1em plus
  0.5em minus 0.4em\relax Yokohama Japan: ACM, May 2021, pp. 1--17. [Online].
  Available: \url{https://dl.acm.org/doi/10.1145/3411764.3445616}
\BIBentrySTDinterwordspacing

\bibitem{kang_detecting_2022}
\BIBentryALTinterwordspacing
H.~J. Kang, K.~L. Aw, and D.~Lo, ``Detecting false alarms from automatic static
  analysis tools: how far are we?'' in \emph{Proceedings of the 44th
  {International} {Conference} on {Software} {Engineering}}, ser. {ICSE}
  '22.\hskip 1em plus 0.5em minus 0.4em\relax New York, NY, USA: Association
  for Computing Machinery, Jul. 2022, pp. 698--709. [Online]. Available:
  \url{https://doi.org/10.1145/3510003.3510214}
\BIBentrySTDinterwordspacing

\bibitem{kim_filtering_2010}
\BIBentryALTinterwordspacing
Y.~Kim, J.~Lee, H.~Han, and K.-M. Choe, ``Filtering false alarms of buffer
  overflow analysis using {SMT} solvers,'' \emph{Information and Software
  Technology}, vol.~52, no.~2, pp. 210--219, Feb. 2010. [Online]. Available:
  \url{https://www.sciencedirect.com/science/article/pii/S095058490900175X}
\BIBentrySTDinterwordspacing

\bibitem{jung_taming_2005}
Y.~Jung, J.~Kim, J.~Shin, and K.~Yi, ``\BIBforeignlanguage{en}{Taming {False}
  {Alarms} from a {Domain}-{Unaware} {C} {Analyzer} by a {Bayesian}
  {Statistical} {Post} {Analysis}},'' in \emph{\BIBforeignlanguage{en}{Static
  {Analysis}}}, ser. Lecture {Notes} in {Computer} {Science}, C.~Hankin and
  I.~Siveroni, Eds.\hskip 1em plus 0.5em minus 0.4em\relax Berlin, Heidelberg:
  Springer, 2005, pp. 203--217.

\bibitem{tomassi_bugs_2018}
\BIBentryALTinterwordspacing
D.~A. Tomassi, ``Bugs in the wild: examining the effectiveness of static
  analyzers at finding real-world bugs,'' in \emph{Proceedings of the 2018 26th
  {ACM} {Joint} {Meeting} on {European} {Software} {Engineering} {Conference}
  and {Symposium} on the {Foundations} of {Software} {Engineering}}, ser.
  {ESEC}/{FSE} 2018.\hskip 1em plus 0.5em minus 0.4em\relax New York, NY, USA:
  Association for Computing Machinery, Oct. 2018, pp. 980--982. [Online].
  Available: \url{https://dl.acm.org/doi/10.1145/3236024.3275439}
\BIBentrySTDinterwordspacing

\bibitem{tomassi_real-world_2021}
D.~A. Tomassi and C.~Rubio-González, ``On the {Real}-{World} {Effectiveness}
  of {Static} {Bug} {Detectors} at {Finding} {Null} {Pointer} {Exceptions},''
  in \emph{2021 36th {IEEE}/{ACM} {International} {Conference} on {Automated}
  {Software} {Engineering} ({ASE})}, Nov. 2021, pp. 292--303, iSSN: 2643-1572.

\bibitem{kavaler_tool_2019}
D.~Kavaler, A.~Trockman, B.~Vasilescu, and V.~Filkov, ``Tool {Choice}
  {Matters}: {JavaScript} {Quality} {Assurance} {Tools} and {Usage} {Outcomes}
  in {GitHub} {Projects},'' in \emph{2019 {IEEE}/{ACM} 41st {International}
  {Conference} on {Software} {Engineering} ({ICSE})}, May 2019, pp. 476--487,
  iSSN: 1558-1225.

\bibitem{tomasdottir_why_2017}
\BIBentryALTinterwordspacing
K.~F. Tomasdottir, M.~Aniche, and A.~van Deursen, ``\BIBforeignlanguage{en}{Why
  and how {JavaScript} developers use linters},'' in
  \emph{\BIBforeignlanguage{en}{2017 32nd {IEEE}/{ACM} {International}
  {Conference} on {Automated} {Software} {Engineering} ({ASE})}}.\hskip 1em
  plus 0.5em minus 0.4em\relax Urbana, IL: IEEE, Oct. 2017, pp. 578--589.
  [Online]. Available: \url{http://ieeexplore.ieee.org/document/8115668/}
\BIBentrySTDinterwordspacing

\bibitem{do_why_2022}
L.~N.~Q. Do, J.~R. Wright, and K.~Ali, ``Why {Do} {Software} {Developers} {Use}
  {Static} {Analysis} {Tools}? {A} {User}-{Centered} {Study} of {Developer}
  {Needs} and {Motivations},'' \emph{IEEE Transactions on Software
  Engineering}, vol.~48, no.~3, pp. 835--847, Mar. 2022, conference Name: IEEE
  Transactions on Software Engineering.

\bibitem{smith_why_2020}
\BIBentryALTinterwordspacing
J.~Smith, L.~N.~Q. Do, and E.~Murphy-Hill, ``\BIBforeignlanguage{en}{Why
  {Can}'t {Johnny} {Fix} {Vulnerabilities}: {A} {Usability} {Evaluation} of
  {Static} {Analysis} {Tools} for {Security}},'' in
  \emph{\BIBforeignlanguage{en}{Sixteenth {Symposium} on {Usable} {Privacy} and
  {Security} ({SOUPS} 2020)}}, 2020, pp. 221--238. [Online]. Available:
  \url{https://www.usenix.org/conference/soups2020/presentation/smith}
\BIBentrySTDinterwordspacing

\bibitem{johnson_cross-tool_2016}
\BIBentryALTinterwordspacing
B.~Johnson, R.~Pandita, J.~Smith, D.~Ford, S.~Elder, E.~Murphy-Hill,
  S.~Heckman, and C.~Sadowski, ``\BIBforeignlanguage{en}{A cross-tool
  communication study on program analysis tool notifications},'' in
  \emph{\BIBforeignlanguage{en}{Proceedings of the 2016 24th {ACM} {SIGSOFT}
  {International} {Symposium} on {Foundations} of {Software}
  {Engineering}}}.\hskip 1em plus 0.5em minus 0.4em\relax Seattle WA USA: ACM,
  Nov. 2016, pp. 73--84. [Online]. Available:
  \url{https://dl.acm.org/doi/10.1145/2950290.2950304}
\BIBentrySTDinterwordspacing

\bibitem{nachtigall_large-scale_2022}
\BIBentryALTinterwordspacing
M.~Nachtigall, M.~Schlichtig, and E.~Bodden, ``\BIBforeignlanguage{en}{A
  large-scale study of usability criteria addressed by static analysis
  tools},'' in \emph{\BIBforeignlanguage{en}{Proceedings of the 31st {ACM}
  {SIGSOFT} {International} {Symposium} on {Software} {Testing} and
  {Analysis}}}.\hskip 1em plus 0.5em minus 0.4em\relax Virtual South Korea:
  ACM, Jul. 2022, pp. 532--543. [Online]. Available:
  \url{https://dl.acm.org/doi/10.1145/3533767.3534374}
\BIBentrySTDinterwordspacing

\bibitem{watson_api_2013}
\BIBentryALTinterwordspacing
R.~Watson, M.~Stamnes, J.~Jeannot-Schroeder, and J.~H. Spyridakis,
  ``\BIBforeignlanguage{en}{{API} documentation and software community values:
  a survey of open-source {API} documentation},'' in
  \emph{\BIBforeignlanguage{en}{Proceedings of the 31st {ACM} international
  conference on {Design} of communication}}.\hskip 1em plus 0.5em minus
  0.4em\relax Greenville North Carolina USA: ACM, Sep. 2013, pp. 165--174.
  [Online]. Available: \url{https://dl.acm.org/doi/10.1145/2507065.2507076}
\BIBentrySTDinterwordspacing

\bibitem{nagappan_empirical_2015}
\BIBentryALTinterwordspacing
M.~Nagappan, R.~Robbes, Y.~Kamei, E.~Tanter, S.~McIntosh, A.~Mockus, and A.~E.
  Hassan, ``\BIBforeignlanguage{en}{An empirical study of goto in {C} code from
  {GitHub} repositories},'' in \emph{\BIBforeignlanguage{en}{Proceedings of the
  2015 10th {Joint} {Meeting} on {Foundations} of {Software}
  {Engineering}}}.\hskip 1em plus 0.5em minus 0.4em\relax Bergamo Italy: ACM,
  Aug. 2015, pp. 404--414. [Online]. Available:
  \url{https://dl.acm.org/doi/10.1145/2786805.2786834}
\BIBentrySTDinterwordspacing

\bibitem{binkley_camelcase_2009}
D.~Binkley, M.~Davis, D.~Lawrie, and C.~Morrell, ``To camelcase or
  under\_score,'' in \emph{2009 {IEEE} 17th {International} {Conference} on
  {Program} {Comprehension}}, May 2009, pp. 158--167, iSSN: 1092-8138.

\bibitem{latappy_replication_2024}
\BIBentryALTinterwordspacing
C.~Latappy, T.~Degueule, J.-R. Falleri, R.~Robbes, X.~Blanc, and C.~Teyton,
  ``\BIBforeignlanguage{en}{Replication {Kit} - {What} the {Fix}? {A} {Study}
  of {ASATs} {Rule} {Documentation}},'' Jan. 2024. [Online]. Available:
  \url{https://zenodo.org/uploads/10522473}
\BIBentrySTDinterwordspacing

\bibitem{zimmermann_card-sorting_2016}
\BIBentryALTinterwordspacing
T.~Zimmermann, ``Card-sorting: {From} text to themes,'' in \emph{Perspectives
  on {Data} {Science} for {Software} {Engineering}}, T.~Menzies, L.~Williams,
  and T.~Zimmermann, Eds.\hskip 1em plus 0.5em minus 0.4em\relax Boston: Morgan
  Kaufmann, Jan. 2016, pp. 137--141. [Online]. Available:
  \url{https://www.sciencedirect.com/science/article/pii/B9780128042069000271}
\BIBentrySTDinterwordspacing

\bibitem{vassallo_how_2020}
\BIBentryALTinterwordspacing
C.~Vassallo, S.~Panichella, F.~Palomba, S.~Proksch, H.~C. Gall, and A.~Zaidman,
  ``\BIBforeignlanguage{en}{How developers engage with static analysis tools in
  different contexts},'' \emph{\BIBforeignlanguage{en}{Empirical Software
  Engineering}}, vol.~25, no.~2, pp. 1419--1457, Mar. 2020. [Online].
  Available: \url{https://doi.org/10.1007/s10664-019-09750-5}
\BIBentrySTDinterwordspacing

\bibitem{passonneau_measuring_2006}
\BIBentryALTinterwordspacing
R.~Passonneau, ``Measuring {Agreement} on {Set}-valued {Items} ({MASI}) for
  {Semantic} and {Pragmatic} {Annotation},'' in \emph{Proceedings of the
  {Fifth} {International} {Conference} on {Language} {Resources} and
  {Evaluation} ({LREC}'06)}.\hskip 1em plus 0.5em minus 0.4em\relax Genoa,
  Italy: European Language Resources Association (ELRA), May 2006. [Online].
  Available:
  \url{http://www.lrec-conf.org/proceedings/lrec2006/pdf/636_pdf.pdf}
\BIBentrySTDinterwordspacing

\bibitem{n_attractive_1984}
\BIBentryALTinterwordspacing
K.~N, ``Attractive {Quality} and {Must}-{Be} {Quality},'' \emph{Journal of the
  Japanese Society for Quality Control}, vol.~31, no.~4, pp. 147--156, 1984.
  [Online]. Available: \url{https://cir.nii.ac.jp/crid/1572261550744179968}
\BIBentrySTDinterwordspacing

\bibitem{begel_analyze_2014}
\BIBentryALTinterwordspacing
A.~Begel and T.~Zimmermann, ``\BIBforeignlanguage{en}{Analyze this! 145
  questions for data scientists in software engineering},'' in
  \emph{\BIBforeignlanguage{en}{Proceedings of the 36th {International}
  {Conference} on {Software} {Engineering}}}.\hskip 1em plus 0.5em minus
  0.4em\relax Hyderabad India: ACM, May 2014, pp. 12--23. [Online]. Available:
  \url{https://dl.acm.org/doi/10.1145/2568225.2568233}
\BIBentrySTDinterwordspacing

\bibitem{guest_applied_2023}
\BIBentryALTinterwordspacing
G.~Guest, K.~M. MacQueen, and E.~E. Namey, ``\BIBforeignlanguage{en}{Applied
  {Thematic} {Analysis}},'' Oct. 2023. [Online]. Available:
  \url{https://uk.sagepub.com/en-gb/eur/applied-thematic-analysis/book233379}
\BIBentrySTDinterwordspacing

\bibitem{dowling_power_2005}
R.~Dowling, ``Power, subjectivity and ethics in qualitative research,'' in
  \emph{Qualitative research methods in human geography}, I.~Hay, Ed.\hskip 1em
  plus 0.5em minus 0.4em\relax South Melbourne, Vic.: Oxford University Press,
  2005, pp. 19--29.

\bibitem{ayewah_using_2008}
\BIBentryALTinterwordspacing
N.~Ayewah, W.~Pugh, D.~Hovemeyer, J.~D. Morgenthaler, and J.~Penix, ``Using
  {Static} {Analysis} to {Find} {Bugs},'' \emph{IEEE Software}, vol.~25, no.~5,
  pp. 22--29, Sep. 2008, conference Name: IEEE Software. [Online]. Available:
  \url{https://ieeexplore.ieee.org/document/4602670}
\BIBentrySTDinterwordspacing

\bibitem{heckman_systematic_2011}
\BIBentryALTinterwordspacing
S.~Heckman and L.~Williams, ``A systematic literature review of actionable
  alert identification techniques for automated static code analysis,''
  \emph{Information and Software Technology}, vol.~53, no.~4, pp. 363--387,
  Apr. 2011. [Online]. Available:
  \url{https://www.sciencedirect.com/science/article/pii/S0950584910002235}
\BIBentrySTDinterwordspacing

\bibitem{johnson_why_2013}
B.~Johnson, Y.~Song, E.~Murphy-Hill, and R.~Bowdidge, ``Why don't software
  developers use static analysis tools to find bugs?'' in \emph{2013 35th
  {International} {Conference} on {Software} {Engineering} ({ICSE})}, May 2013,
  pp. 672--681, iSSN: 1558-1225.

\bibitem{gorski_listen_2020}
\BIBentryALTinterwordspacing
P.~L. Gorski, Y.~Acar, L.~Lo~Iacono, and S.~Fahl,
  ``\BIBforeignlanguage{en}{Listen to {Developers}! {A} {Participatory}
  {Design} {Study} on {Security} {Warnings} for {Cryptographic} {APIs}},'' in
  \emph{\BIBforeignlanguage{en}{Proceedings of the 2020 {CHI} {Conference} on
  {Human} {Factors} in {Computing} {Systems}}}.\hskip 1em plus 0.5em minus
  0.4em\relax Honolulu HI USA: ACM, Apr. 2020, pp. 1--13. [Online]. Available:
  \url{https://dl.acm.org/doi/10.1145/3313831.3376142}
\BIBentrySTDinterwordspacing

\bibitem{buckers_uav_2017}
T.~Buckers, C.~Cao, M.~Doesburg, B.~Gong, S.~Wang, M.~Beller, and A.~Zaidman,
  ``{UAV}: {Warnings} from multiple {Automated} {Static} {Analysis} {Tools} at
  a glance,'' in \emph{2017 {IEEE} 24th {International} {Conference} on
  {Software} {Analysis}, {Evolution} and {Reengineering} ({SANER})}, Feb. 2017,
  pp. 472--476.

\bibitem{monperrus_what_2012}
\BIBentryALTinterwordspacing
M.~Monperrus, M.~Eichberg, E.~Tekes, and M.~Mezini, ``What {Should}
  {Developers} {Be} {Aware} {Of}? {An} {Empirical} {Study} on the {Directives}
  of {API} {Documentation},'' \emph{Empirical Software Engineering}, vol.~17,
  no.~6, pp. 703--737, 2012. [Online]. Available:
  \url{http://www.monperrus.net/martin/An-Empirical-Study-On-the-Directives-of-API-Documentation.pdf}
\BIBentrySTDinterwordspacing

\bibitem{cummaudo_what_2019}
A.~Cummaudo, R.~Vasa, and J.~Grundy, ``What should {I} document? {A}
  preliminary systematic mapping study into {API} documentation knowledge,'' in
  \emph{2019 {ACM}/{IEEE} {International} {Symposium} on {Empirical} {Software}
  {Engineering} and {Measurement} ({ESEM})}, Sep. 2019, pp. 1--6, iSSN:
  1949-3789.

\bibitem{aghajani_software_2019}
\BIBentryALTinterwordspacing
E.~Aghajani, C.~Nagy, O.~L. Vega-Márquez, M.~Linares-Vásquez, L.~Moreno,
  G.~Bavota, and M.~Lanza, ``Software {Documentation} {Issues} {Unveiled},'' in
  \emph{2019 {IEEE}/{ACM} 41st {International} {Conference} on {Software}
  {Engineering} ({ICSE})}, May 2019, pp. 1199--1210, iSSN: 1558-1225. [Online].
  Available: \url{https://ieeexplore.ieee.org/abstract/document/8811931}
\BIBentrySTDinterwordspacing

\bibitem{aghajani_software_2020}
\BIBentryALTinterwordspacing
E.~Aghajani, C.~Nagy, M.~Linares-Vásquez, L.~Moreno, G.~Bavota, M.~Lanza, and
  D.~C. Shepherd, ``Software documentation: the practitioners' perspective,''
  in \emph{Proceedings of the {ACM}/{IEEE} 42nd {International} {Conference} on
  {Software} {Engineering}}, ser. {ICSE} '20.\hskip 1em plus 0.5em minus
  0.4em\relax New York, NY, USA: Association for Computing Machinery, Oct.
  2020, pp. 590--601. [Online]. Available:
  \url{https://doi.org/10.1145/3377811.3380405}
\BIBentrySTDinterwordspacing

\bibitem{forward_relevance_2002}
\BIBentryALTinterwordspacing
A.~Forward and T.~C. Lethbridge, ``The {Relevance} of {Software}
  {Documentation}, {Tools} and {Technologies}: {A} {Survey},'' in
  \emph{Proceedings of the 2002 {ACM} {Symposium} on {Document} {Engineering}},
  ser. {DocEng} '02.\hskip 1em plus 0.5em minus 0.4em\relax New York, NY, USA:
  ACM, 2002, pp. 26--33. [Online]. Available:
  \url{http://doi.acm.org/10.1145/585058.585065}
\BIBentrySTDinterwordspacing

\bibitem{roehm2012professional}
T.~Roehm, R.~Tiarks, R.~Koschke, and W.~Maalej, ``How do professional
  developers comprehend software?'' in \emph{2012 34th International Conference
  on Software Engineering (ICSE)}.\hskip 1em plus 0.5em minus 0.4em\relax IEEE,
  2012, pp. 255--265.

\end{thebibliography}

\end{document}